
%
%
%
\catcode`@=11
\expandafter\ifx\csname inp@t\endcsname\relax\let\inp@t=\input
\def\input#1 {\expandafter\ifx\csname #1IsLoaded\endcsname\relax
\inp@t#1%
\expandafter\def\csname #1IsLoaded\endcsname{(#1 was previously loaded)}
\else\message{\csname #1IsLoaded\endcsname}\fi}\fi
\catcode`@=12

\font\ninerm=cmr9    \font\ninei=cmmi9
\font\ninesy=cmsy9 
\font\ninebf=cmbx9 \font\ninesl=cmsl9
\font\ninett=cmtt9 \font\nineit=cmti9
\skewchar\ninei='177   \skewchar\ninesy='60
\hyphenchar\ninett=-1
 \font\eighti=cmmi8
\font\eightsy=cmsy8 
\font\eighttt=cmtt8 

\skewchar\eighti='177 \skewchar\eightsy='60
\hyphenchar\eighttt=-1
\font\sixrm=cmr6 \font\sixi=cmmi6
\font\sixsy=cmsy6 \font\sixbf=cmbx6
\skewchar\sixi='177 \skewchar\sixsy='60
\hyphenchar\tentt=-1


\def\ninepoint{\normalbaselineskip=9.5pt plus 0.1pt minus 0.1pt
  \abovedisplayskip 9.5pt plus 3pt minus 8pt
  \belowdisplayskip 9.5pt plus 3pt minus 8pt
  \abovedisplayshortskip 0pt plus 3pt
  \belowdisplayshortskip 5.4pt plus 3pt minus 4pt
  \smallskipamount=3pt plus1pt minus1pt
  \medskipamount=5.4pt plus2pt minus2pt
  \bigskipamount=9.5pt plus4pt minus4pt
  \def\rm{\fam0\ninerm}          \def\it{\fam\itfam\nineit}%
  \def\sl{\fam\slfam\ninesl}     \def\bf{\fam\bffam\ninebf}%
  \def\sc{\ninebf}             \def\mit{\fam 1}%
  \def\cal{\fam 2}%
  \def\tt{\ninett}             \def\sf{niness}
 \textfont0=\ninerm \scriptfont0=\sixrm \scriptscriptfont0=\fiverm
 \textfont1=\ninei \scriptfont1=\sixi \scriptscriptfont1=\fivei
 \textfont2=\ninesy \scriptfont2=\sixsy \scriptscriptfont2=\fivesy
 \textfont3=\tenex \scriptfont3=\tenex \scriptscriptfont3=\tenex
 \textfont\itfam=\nineit
 \textfont\slfam=\ninesl
 \textfont\ttfam=\ninett
 \textfont\bffam=\ninebf \scriptfont\bffam=\sixbf
  \scriptscriptfont\bffam=\fivebf
  \normalbaselines\rm}

\def\beginlinemode{\endmode
  \begingroup\parskip=0pt \obeylines\def\\{\par}\def\endmode{\par\endgroup}}
\def\beginparmode{\endmode
  \begingroup \def\endmode{\par\endgroup}}
\let\endmode=\par
{\obeylines\gdef\
{}}
\def\singlespace{\baselineskip=\normalbaselineskip}
\def\doublespace{\baselineskip=\normalbaselineskip \multiply\baselineskip by 2}
\def\oneandathirdspace{\baselineskip=\normalbaselineskip
  \multiply\baselineskip by 4 \divide\baselineskip by 3}
\def\oneandahalfspace{\baselineskip=\normalbaselineskip
  \multiply\baselineskip by 3 \divide\baselineskip by 2}

\newcount\firstpageno
\firstpageno=2
\footline={\ifnum\pageno<\firstpageno{\hfil}\else{\hfil\ninerm\folio\hfil}\fi}
\def\toppageno{\global\footline={\hfil}\global\headline
  ={\ifnum\pageno<\firstpageno{\hfil}\else{\hfil\ninerm\folio\hfil}\fi}}
\let\rawfootnote=\footnote		
\def\footnote#1#2{{\rm\singlespace\parindent=0pt\parskip=0pt
  \rawfootnote{#1}{#2\hfill\vrule height 0pt depth 6pt width 0pt}}}
\def\raggedcenter{\leftskip=4em plus 12em \rightskip=\leftskip
  \parindent=0pt \parfillskip=0pt \spaceskip=.3333em \xspaceskip=.5em
  \pretolerance=9999 \tolerance=9999
  \hyphenpenalty=9999 \exhyphenpenalty=9999 }

\parskip=\medskipamount
\def\\{\cr}
\ninepoint		
\overfullrule=0pt	
\def\title			
  {\null\vskip 3pt plus 0.2fill
   \beginlinemode \doublespace \raggedcenter \bf}

\def\author			
  {\vskip 3pt plus 0.2fill \beginlinemode
   \singlespace \raggedcenter\sc}

\def\affil			
  {\vskip 3pt plus 0.1fill \beginlinemode
   \oneandahalfspace \raggedcenter \sl}

\def\endtitlepage		
  {\endpage			
   \body}

\def\body			
  {\beginparmode}		

\def\head#1{			
  \goodbreak\vskip 0.5truein	
  {\immediate\write16{#1}
   \raggedcenter \uppercase{#1}\par}
   \nobreak\vskip 0.25truein\nobreak}

\def\subhead#1{			
  \vskip 0.25truein		
  {\raggedcenter {#1} \par}
   \nobreak\vskip 0.25truein\nobreak}

\def\beginitems{
\par\medskip\bgroup\def\i##1 {\item{##1}}\def\ii##1 {\itemitem{##1}}
\leftskip=36pt\parskip=0pt}
\def\enditems{\par\egroup}

\def\beneathrel#1\under#2{\mathrel{\mathop{#2}\limits_{#1}}}

\def\references			
  {\head{References}		
   \beginparmode
   \frenchspacing \parindent=0pt \leftskip=1truecm
   \parskip=8pt plus 3pt \everypar{\hangindent=\parindent}}

\gdef\refis#1{\item{#1.\ }}			

\gdef\journal#1, #2, #3, 1#4#5#6{		
    {\sl #1~}{\bf #2}, #3 (1#4#5#6)}		

\def\pr{\journal Phys. Rev., }

\def\endreferences{\body}

\def\endpage			
  {\vfill\eject}

\def\endpaper			
  {\endmode\vfill\supereject}
\def\endit
  {\endpaper\end}

\def\cite#1{{#1}}
\def\(#1){(\call{#1})}
\def\call#1{{#1}}
\def\taghead#1{}
\def\12{{1\over2}}
\def\ie{{\it i.e.,\ }}

\newdimen\fullhsize
\fullhsize=9.5truein
\def\fullline{\hbox to\fullhsize}
\hsize=4.5truein
\hoffset=0.truein
\vsize=7.00truein
\voffset=-0.13truein
\oneandathirdspace

\def\abstract			
  {\vskip 3pt plus 0.3fill \beginparmode
   \singlespace ABSTRACT: }

\let\lr=L \newbox\leftcolumn
\output={\if L\lr
    \global\setbox\leftcolumn=\columnbox \global\let\lr=R
  \else \doubleformat \global\let\lr=L\fi
  \advancepageno
  \ifnum\outputpenalty>-20000 \else\dosupereject\fi}
\def\doubleformat{\shipout\vbox{\fullline{\box\leftcolumn\hfil\columnbox}}
}
\def\columnbox{\leftline{\vbox{\makeheadline\pagebody\makefootline}}}

\def\endpaper{\endmode\vfill\supereject\if R\lr \null\vfill\eject\fi
  \immediate\write16{>>>>To print this, use the command
  IMPRINT/MAG=1250/LAND \jobname.dvi<<<<}}
\catcode`@=11
\newcount\r@fcount \r@fcount=0
\newcount\r@fcurr
\immediate\newwrite\reffile
\newif\ifr@ffile\r@ffilefalse
\def\w@rnwrite#1{\ifr@ffile\immediate\write\reffile{#1}\fi\message{#1}}

\def\writer@f#1>>{}
\def\referencefile{
  \r@ffiletrue\immediate\openout\reffile=\jobname.ref%
  \def\writer@f##1>>{\ifr@ffile\immediate\write\reffile%
    {\noexpand\refis{##1} = \csname r@fnum##1\endcsname = %
     \expandafter\expandafter\expandafter\strip@t\expandafter%
     \meaning\csname r@ftext\csname r@fnum##1\endcsname\endcsname}\fi}%
  \def\strip@t##1>>{}}

\def\citeall#1{\xdef#1##1{#1{\noexpand\cite{##1}}}}
\def\cite#1{\each@rg\citer@nge{#1}}	

\def\each@rg#1#2{{\let\thecsname=#1\expandafter\first@rg#2,\end,}}
\def\first@rg#1,{\thecsname{#1}\apply@rg}	
\def\apply@rg#1,{\ifx\end#1\let\next=\relax
\else,\thecsname{#1}\let\next=\apply@rg\fi\next}

\def\citer@nge#1{\citedor@nge#1-\end-}	
\def\citer@ngeat#1\end-{#1}
\def\citedor@nge#1-#2-{\ifx\end#2\r@featspace#1 
  \else\citel@@p{#1}{#2}\citer@ngeat\fi}	
\def\citel@@p#1#2{\ifnum#1>#2{\errmessage{Reference range #1-#2\space is bad.}%
    \errhelp{If you cite a series of references by the notation M-N, then M and
    N must be integers, and N must be greater than or equal to M.}}\else%
 {\count0=#1\count1=#2\advance\count1
by1\relax\expandafter\r@fcite\the\count0,%
  \loop\advance\count0 by1\relax
    \ifnum\count0<\count1,\expandafter\r@fcite\the\count0,%
  \repeat}\fi}

\def\r@featspace#1#2 {\r@fcite#1#2,}	
\def\r@fcite#1,{\ifuncit@d{#1}
    \newr@f{#1}%
    \expandafter\gdef\csname r@ftext\number\r@fcount\endcsname%
                     {\message{Reference #1 to be supplied.}%
                      \writer@f#1>>#1 to be supplied.\par}%
 \fi%
 \csname r@fnum#1\endcsname}
\def\ifuncit@d#1{\expandafter\ifx\csname r@fnum#1\endcsname\relax}%
\def\newr@f#1{\global\advance\r@fcount by1%
    \expandafter\xdef\csname r@fnum#1\endcsname{\number\r@fcount}}

\let\r@fis=\refis			
\def\refis#1#2#3\par{\ifuncit@d{#1}
   \newr@f{#1}%
   \w@rnwrite{Reference #1=\number\r@fcount\space is not cited up to now.}\fi%
  \expandafter\gdef\csname r@ftext\csname r@fnum#1\endcsname\endcsname%
  {\writer@f#1>>#2#3\par}}

\def\ignoreuncited{
   \def\refis##1##2##3\par{\ifuncit@d{##1}%
     \else\expandafter\gdef\csname r@ftext\csname
r@fnum##1\endcsname\endcsname%
     {\writer@f##1>>##2##3\par}\fi}}

\def\r@ferr{\endreferences\errmessage{I was expecting to see
\noexpand\endreferences before now;  I have inserted it here.}}
\let\r@ferences=\references
\def\references{\r@ferences\def\endmode{\r@ferr\par\endgroup}}

\let\endr@ferences=\endreferences
\def\endreferences{\r@fcurr=0
  {\loop\ifnum\r@fcurr<\r@fcount
    \advance\r@fcurr by 1\relax\expandafter\r@fis\expandafter{\number\r@fcurr}%
    \csname r@ftext\number\r@fcurr\endcsname%
  \repeat}\gdef\r@ferr{}\endr@ferences}

\let\r@fend=\endpaper\gdef\endpaper{\ifr@ffile
\immediate\write16{Cross References written on []\jobname.REF.}\fi\r@fend}

\catcode`@=12

\catcode`@=11
\newcount\tagnumber\tagnumber=0

\immediate\newwrite\eqnfile
\newif\if@qnfile\@qnfilefalse
\def\write@qn#1{}
\def\writenew@qn#1{}
\def\w@rnwrite#1{\write@qn{#1}\message{#1}}
\def\@rrwrite#1{\write@qn{#1}\errmessage{#1}}

\def\taghead#1{\gdef\t@ghead{#1}\global\tagnumber=0}
\def\t@ghead{}

\expandafter\def\csname @qnnum-3\endcsname
  {{\t@ghead\advance\tagnumber by -3\relax\number\tagnumber}}
\expandafter\def\csname @qnnum-2\endcsname
  {{\t@ghead\advance\tagnumber by -2\relax\number\tagnumber}}
\expandafter\def\csname @qnnum-1\endcsname
  {{\t@ghead\advance\tagnumber by -1\relax\number\tagnumber}}
\expandafter\def\csname @qnnum0\endcsname
  {\t@ghead\number\tagnumber}
\expandafter\def\csname @qnnum+1\endcsname
  {{\t@ghead\advance\tagnumber by 1\relax\number\tagnumber}}
\expandafter\def\csname @qnnum+2\endcsname
  {{\t@ghead\advance\tagnumber by 2\relax\number\tagnumber}}
\expandafter\def\csname @qnnum+3\endcsname
  {{\t@ghead\advance\tagnumber by 3\relax\number\tagnumber}}

\def\equationfile{%
  \@qnfiletrue\immediate\openout\eqnfile=\jobname.eqn%
  \def\write@qn##1{\if@qnfile\immediate\write\eqnfile{##1}\fi}
  \def\writenew@qn##1{\if@qnfile\immediate\write\eqnfile
    {\noexpand\tag{##1} = (\t@ghead\number\tagnumber)}\fi}
}

\def\callall#1{\xdef#1##1{#1{\noexpand\call{##1}}}}
\def\call#1{\each@rg\callr@nge{#1}}

\def\each@rg#1#2{{\let\thecsname=#1\expandafter\first@rg#2,\end,}}
\def\first@rg#1,{\thecsname{#1}\apply@rg}
\def\apply@rg#1,{\ifx\end#1\let\next=\relax%
\else,\thecsname{#1}\let\next=\apply@rg\fi\next}

\def\callr@nge#1{\calldor@nge#1-\end-}
\def\callr@ngeat#1\end-{#1}
\def\calldor@nge#1-#2-{\ifx\end#2\@qneatspace#1 %
  \else\calll@@p{#1}{#2}\callr@ngeat\fi}
\def\calll@@p#1#2{\ifnum#1>#2{\@rrwrite{Equation range #1-#2\space is bad.}
\errhelp{If you call a series of equations by the notation M-N, then M and
N must be integers, and N must be greater than or equal to M.}}\else%
 {\count0=#1\count1=#2\advance\count1
by1\relax\expandafter\@qncall\the\count0,%
  \loop\advance\count0 by1\relax%
    \ifnum\count0<\count1,\expandafter\@qncall\the\count0,%
  \repeat}\fi}

\def\@qneatspace#1#2 {\@qncall#1#2,}
\def\@qncall#1,{\ifunc@lled{#1}{\def\next{#1}\ifx\next\empty\else
  \w@rnwrite{Equation number \noexpand\(>>#1<<) has not been defined yet.}
  >>#1<<\fi}\else\csname @qnnum#1\endcsname\fi}

\let\eqnono=\eqno
\def\eqno(#1){\tag#1}
\def\tag#1$${\eqnono(\displayt@g#1 )$$}

\def\aligntag#1\endaligntag
  $${\gdef\tag##1\\{&(##1 )\cr}\eqalignno{#1\\}$$
  \gdef\tag##1$${\eqnono(\displayt@g##1 )$$}}

\def\eqalignno#1{\displ@y \tabskip\centering
  \halign to\displaywidth{\hfil$\displaystyle{##}$\tabskip\z@skip
    &$\displaystyle{{}##}$\hfil\tabskip\centering
    &\llap{$\displayt@gpar##$}\tabskip\z@skip\crcr
    #1\crcr}}

\def\displayt@gpar(#1){(\displayt@g#1 )}

\def\displayt@g#1 {\rm\ifunc@lled{#1}\global\advance\tagnumber by1
        {\def\next{#1}\ifx\next\empty\else\expandafter
        \xdef\csname @qnnum#1\endcsname{\t@ghead\number\tagnumber}\fi}%
  \writenew@qn{#1}\t@ghead\number\tagnumber\else
        {\edef\next{\t@ghead\number\tagnumber}%
        \expandafter\ifx\csname @qnnum#1\endcsname\next\else
        \w@rnwrite{Equation \noexpand\tag{#1} is a duplicate number.}\fi}%
  \csname @qnnum#1\endcsname\fi}

\def\ifunc@lled#1{\expandafter\ifx\csname @qnnum#1\endcsname\relax}

\let\@qnend=\end\gdef\end{\if@qnfile
\immediate\write16{Equation numbers written on []\jobname.EQN.}\fi\@qnend}

\catcode`@=12

\def\ssc{\scriptscriptstyle}
\def\scc{\scriptstyle}
\def\-#1{_{\ssc {#1} }}
\def\s #1{{\cal {#1}}}
\def\ie{{\it i.e.,\ }}

\def\hat #1{\mathaccent94{#1}}
\def\sgn#1{\,\hbox{sgn}{\scc (}{#1}{\scc )}\,}
\def\D{{\s D}}
\def\lint {\int_{-\Lambda}^\Lambda\!}

\def\alphabar {\bar\alpha}
\def\b #1{\sqrt{|{#1}|\pi}}
\hsize=4.5truein
\vsize=6.5truein

\title Influence~functionals and black~body~radiation

\author J.R. Anglin\footnote{$^\dagger$}{anglin@hep.physics.mcgill.ca}

\affil Physics Department, McGill University
3600 University Street
Montr\'eal, Qu\'ebec CANADA H3A 2T8

\abstract  The Feynman-Vernon influence functional formalism is used to derive
the effect of a black body, treated as an environment, on a massless scalar
field in 1+1 dimensions.  The black body is modeled as a finite region of space
in which an independent ohmic heat bath is weakly coupled to the field at every
point.  A weak coupling approximation is  developed which implements the
concept of an ideal black body in the context of quantum field theory.  The
calculation takes advantage of the suitability to harmonic oscillators and free
fields of  Bargmann-Fock coherent state variables, whose convenience is
illustrated by a preliminary derivation of the thermalization of a single
harmonic oscillator by a heat bath with slowly varying spectral density.  The
black body model exhibits absorption, thermal equilibrium, and emission
consistent with classical results for black bodies. It is argued that this
model is in fact a realistic description of a very fine, granular medium, such
as lampblack.

\endtitlepage

\eject

\head{\bf I.  Introduction}
Historically, quantum theory began with black body radiation[\cite{Planck}].
In retrospect, black body radiation is the kind of phenomenon in which one
might have expected to see new physics.  The Maxwell-Lorentz theory provided a
well-defined theoretical framework for electromagnetic radiation interacting
with matter, and statistical mechanics provided a model for thermal
equilibrium; but the two theories had essentially been developed independently
of each other, and so their joint application to a single problem was a
non-trivial test of the two ideas that stood as the principal achievements
of physics in the nineteenth century[\cite{RayleighJeans}].  And, in the form
of
the Hawking effect, a very similar place at the crux of a century's major
physical theories is still held by black body radiation[\cite{Jacobsen}].

Pedagogically, quantum theory typically begins with a quantum system coupled to
an external source, treated classically.  This source is meant to represent
some sort of environment, not itself directly under investigation, which is
coupled to the system under observation.  One effectively ignores the
environmental degrees of freedom, and replaces the environmental part of the
interaction Hamiltonian with its expectation value.  A thorough understanding
of the role of classical concepts in quantum mechanics is still being sought,
of course[\cite{deco, deco2}], but part of the explanation for the successes of
external source models is that this external source approximation is accurate
to first order in the coupling strength, and in the most common case of
electromagnetic interactions, this coupling is weak enough that higher order
effects may often be neglected.

There are some important situations, however, in which the quantum fluctuations
around the expectation value of the coupling operator produce effects on the
observed system that are not small, despite being of second order in a small
coupling constant $g$.  This happens because other parameters enter the
problem, which are large enough to outweigh the smallness of $g^2$.  In cases
such as these there may be two kinds of second order terms --- non-negligible
terms containing the large parameter, as well as negligible contributions which
do not.  For example, when a previously uncorrelated system and environment are
suddenly made to interact, the second order effects include quantum
measurement, in which the rate at which the wave function collapses is set by
the high ultraviolet cut-off of the interaction with the unobserved
environment[\cite{wfcollapse}].  On the other hand, a sufficiently long time
scale can also enable the second order terms to drive a system into
thermodynamic equilibrium with its environment[\cite{Grabert, DvH}].  An
analogous phenomenon will be shown in this paper to explain the fact that the
thermal radiation of a black body, although it is generated via the coupling of
photons to charged matter, is not proportional in intensity to the weak
electromagnetic  coupling\footnote{$^\dagger$}{It is a fact that should be more
surprising than it is, that the Stefan-Boltzmann constant  $\sigma = {2\pi^5
k\-B^4\over 15 h^3 c^2}$ is independent of the elementary charge $e$.}.

Second order environmental effects of these kinds are not described by
classical external sources, but may be accurately expressed in the influence
functional formalism of Feynman and Vernon[\cite{FV}].  This  formalism has
already been used to examine quantum  measurement[\cite{wfcollapse}].  In the
present paper, it  is applied to radiation and absorption by an ideal black
body of finite extent.  Inasmuch as we are able to describe radiation
propagating away from a heat source, our results go beyond the standard
equilibrium treatments.

This paper is organized as follows.  The problem of a simple harmonic
oscillator driven to thermal equilibrium by contact (velocity coupling) with a
heat bath is presented first, in the comparatively little-used formalism of the
path integral in Bargmann-Fock coherent state variables.  Although these
variables are less familiar than positions and momenta, they are actually
particularly well-suited to problems involving harmonic oscillators, including
any calculations involving free quantum fields.  They are introduced here in a
demonstration that standard thermal equilibrium is established, after a
sufficiently long time, whenever the coupling to the bath is weak enough and
its spectral density  varies slowly enough; a  spectral density obeying a
particular ``ohmic'' power law is not required.  For simplicity, we then assume
an ohmic environment with constant spectral density, and consider a  massless
quantum field coupled to a heat bath restricted to a finite region of space.
Here the coherent state variables are particularly advantageous, allowing us to
clearly see that this model exhibits uniform absorption and black body
radiation.  We then conclude with a discussion of the realism of the model
chosen, the issues raised by our analysis, and some related problems for future
study.  An appendix reviews the important justification, originally given by
Feynman and Vernon[\cite{FV}], for treating a generic weakly-coupled
environment as a bath of independent harmonic oscillators.
\vfill

\head{\bf II. Oscillator with a heat bath in Bargmann-Fock variables}
\subhead{\bf A. The Bargmann-Fock path integral}
When using path integrals to obtain the time evolution of a quadratic  quantum
mechanical system (\ie one whose action has canonical kinetic terms,  quadratic
potentials, and bilinear interactions), the problem may be solved in two
stages.  First, one solves an equation of motion with initial and final
conditions; and second, one averages over the initial and final conditions
using initial and final wave functions as weights.  These averages may be easy
to compute, if the initial and final states have particularly simple wave
functions in the representation in which the path integral is constructed.  For
example, in the usual position representation, position eigenstates are
described by delta functions.  It  turns out, however, that in the black body
problem considered below, describing the evolution of $N$-particle states in
field variables leads to some integrals over initial data that are difficult to
perform, and whose results are difficult to interpret.  (These integrals are
generalized Gaussians, whose exponents involve non-local kernels that are
difficult to invert because of the finite spatial extent of the black body
environment.)   Constructing the path integral in a different set of variables,
in which $N$-particle states have simple representations, avoids this obstacle,
and also makes the final answer quite transparent.  These Bargmann-Fock
variables will be introduced here, in the simpler problem of a single harmonic
oscillator coupled to a heat bath, so that this toy model will provide an
indication of both the physics and the formalism of the main problem.  The
derivation of the path integral in Bargmann-Fock space  is described in a
standard text[\cite{ItZub}], but it will be reviewed briefly here.

To span a Hilbert space, one chooses bases for both vectors and dual vectors.
Often, one chooses the duals of the ket basis elements to form the basis of
bras, but in general this need not be the case.  For example, one may take
position states as a basis for kets, and momentum states as a basis for bras.
These bases are convenient for free particles, but for a harmonic oscillator,
there are better alternatives, namely bases of eigenstates of the annihilation
operator (coherent states).

Consider the following coherent states as basis elements for bras and kets:
$$
|\alpha\rangle = \sum_{n=0}^\infty {\alpha^n\over\sqrt{n!}}
|n\rangle \qquad\qquad
\langle\bar\alpha | = \sum_{n=0}^\infty {\bar\alpha^n\over\sqrt{n!}}\langle n|
\;\;,\eqno(basis)
$$
where the $|n\rangle$ are harmonic oscillator energy eigenstates, and $\alpha$
and $\bar\alpha$ are complex numbers (not necessarily complex conjugates
of each other).  Since a whole 2-plane of parameter space is larger than needed
to span the states of an oscillator, we constrain $\alpha$ to lie on a line $C$
through the origin, and $\bar\alpha$ to lie on another line $\bar C$ through
the origin, such that $\bar C$ is perpendicular to $C$.  We then have an
expression for the identity operator,
$$
{1\over2\pi i}\int_C\!d\alpha\int_{\bar C}\!d\bar\alpha\, e^{-\bar\alpha\alpha}
\,|\alpha\rangle\langle\bar\alpha |\; = 1\;.\eqno(iden)
$$
(This relation may be verified by inserting the definitions \(basis) and using
differentiation under the integral sign, treating the expression as a
distribution.)

Since $|\alpha\rangle$ is an eigenstate of the annihilation operator $\hat a$,
any operator expressed as a normal-ordered function of creation and
annihilation operators has a simple matrix form in this representation:
$$
\langle\bar\alpha |:\!A(\hat a^\dagger,\hat a)\!:|\alpha\rangle =
A(\bar\alpha,\alpha)\;.\eqno(Arep)
$$
In particular, for a Hamiltonian $\hat H=\,:\!\!H(\hat a^\dagger,\hat a)\!\!:$,
we  have the following matrix elements of the infinitesimal time evolution
operator:
$$
\langle\bar\alpha |e^{-i\delta t\, \hat H}|\alpha\rangle = e^{-i\delta t\,
h(\bar\alpha,\alpha)} + {\s O}(\delta t^2)\;.\eqno(eiht)
$$
(Throughout this paper we set $\hbar=c=1$.)  Using \(iden) and \(eiht), we can
derive the path integral for a transition amplitude in the usual way, by
inserting identity operators between an infinite succession of infinitesimal
time intervals.  We obtain the Bargmann-Fock path integral
$$\eqalign{
\langle\bar\alpha\-f |e^{-i\int_0^t\!ds\,\hat H(s)}&|\alpha\-i\rangle =
\int\!{\s D}\alpha{\s D}\bar\alpha\, \delta\bigl(\alpha(0) - \alpha\-i\bigr)
\delta\bigl(\bar\alpha(t) - \bar\alpha\-f\bigr)\cr
&\times \exp{\left[ {{\bar\alpha\-f\alpha(t) + \bar\alpha(0)\alpha\-i}\over2}
  +i\int_0^t\!ds\,\left({{\dot{\bar\alpha}\alpha -
   \bar\alpha\dot{\alpha}}\over2i}
   - H(\bar\alpha,\alpha)\right)\right]}\;,}\eqno(bfpi)
$$
$s$ being used as a time parameter, with $\dot\alpha \equiv {d\ \over ds}
\alpha(s)$.  Note that, according to the delta functions in \(bfpi),
$\alpha(s)$ and $\bar\alpha(s)$ have only one boundary condition each, at the
initial time $s=0$, and the final time $s=t$, respectively.

In the case where the Hamiltonian $H$ is quadratic in $\alpha$ and
$\bar\alpha$,   \(bfpi) is equal, up to a normalization factor, to the value of
the integrand when its exponent is extremized.  As usual, this condition is
satisfied by  imposing the Euler-Lagrange equations.  In this case, the
Euler-Lagrange equations are first order, and we find that only the boundary
terms do not vanish on shell.  The transition amplitude therefore simplifies to
$$
\langle\bar\alpha\-f |e^{-i\int_0^t\!ds\,H(s)}|\alpha\-i\rangle = {1\over Z}
e^{{\scc{1\over2}} \bigl(\bar\alpha\-f\alpha(t) +
\bar\alpha(0)\alpha\-i\bigr)\big\vert_0}
\qquad\hbox{[quadratic $h$]}\;,\eqno(linh)
$$
for some normalization factor $Z$.  The subscript $0$ indicates that
$\alpha(t)$ and $\alphabar(0)$ are endpoint values of the solutions of the
Euler-Lagrange equations.  For example, in the case of a harmonic oscillator
driven by an external source, with
$$
\hat H(s) = \Omega (\hat a^\dagger\hat a + \12) +
f(s)\hat a^\dagger + \bar f(s)\hat a\;,\eqno(hatHs)
$$
where $f$ and $\bar f$ are c-number functions representing the time-dependent
source, these solutions may be obtained by direct integration, without the need
for Green's functions.  The resulting endpoint  values will then depend
linearly on the driving sources, as well as on the initial and final
conditions:
$$\eqalign{
\alpha(t) &= \alpha\-i e^{-i\Omega t} +
                           i\int_0^t\! ds\, e^{-i\Omega(t-s)}f(s)\cr
\alphabar(0) &= \alphabar\-f e^{-i\Omega t} +
                           i\int_0^t\!ds\, e^{-i\Omega s} \bar f(s)\;.}
\eqno(endpoint)
$$
While this paper deals with interactions that cannot be represented by simple
external sources of this type, similar calculations involving quadratic
actions, Euler-Lagrange equations, and boundary values will be performed
repeatedly below.

Because $\langle n|\alpha\rangle = {\alpha^n\over\sqrt{n!}}$, it is easy to
obtain transition amplitudes between $|n\rangle$ states, without integrating,
from \(linh).  One simply notices that
$$
\langle\bar\alpha\-f |e^{-i\int_0^t\!ds\,\hat H(s)}|\alpha\-i\rangle =
 \sum_{m,n=0}^\infty {\bar\alpha\-f^m\alpha\-i^n\over\sqrt{m!n!}}
 \langle m|e^{-i\int_0^t\!ds\,\hat H(s)}|n\rangle\;,\eqno(mnamp)
$$
and extracts the co-efficients of the powers of $\alpha\-i$ and
$\bar\alpha\-f$.  In the case of the driven oscillator, this gives a clear
interpretation of the extremum endpoint values in \(endpoint).   The terms in
$\alpha(t)$   proportional to $\alpha\-i$, and in $\bar\alpha(0)$ proportional
to $\bar\alpha\-f$, describe energy-conserving evolution of initial states into
final states.  The other term in $\alpha(t)$ describes excitation of the
system, by the driving source, while the other term in $\bar\alpha(0)$ implies
absorption by the source.  This interpretation is emphasized here in order to
provide some intuition for the results of Section III, in which a quantum
mechanical black body will appear instead of a classical driving force, but in
which the action will still be quadratic.  All of the system's dynamics will
still be expressed in the boundary values of $\alpha$ and $\bar\alpha$
variables.

The basis in which a path integral is expressed determines the type of boundary
conditions which the paths must satisfy.  For given boundary conditions,
though, one is free to change variables in the path integral.  For example, we
are free to re-express \(bfpi) as
$$\eqalign{
\langle&\bar\alpha\-f |e^{-i\int_0^t\!ds\,\hat H(s)}|\alpha\-i\rangle =\cr
&\int\!{\s D}P{\s D}Q\, \delta\bigl(\Omega Q(0) +iP(0) - \sqrt{2\Omega}
    \alpha\-i\bigr)
\delta\bigl(\Omega Q(t) -iP(t)  - \sqrt{2\Omega}\bar\alpha\-f\bigr)\cr
&\qquad\qquad\times\exp\Bigl[ {{\bar\alpha\-f[\Omega Q(t)+iP(t)]
            + \alpha\-i[\Omega Q(0)-iP(0)]}  \over2\sqrt{2\Omega}}\cr
&\qquad\qquad\qquad  +i\int_0^t\!ds\,\left({{P\dot{Q} - Q\dot{P}}\over2}
              - h(P,Q)\right)\Bigr]\;,}\eqno(bfPQ)
$$
using the transformation
$$\alpha = {1\over\sqrt{2\Omega}}(\Omega Q + iP)\qquad\qquad\qquad
\bar\alpha = {1\over\sqrt{2\Omega}}(\Omega Q - iP) \eqno(QP)
$$
and absorbing the constant Jacobian in the new measure.  While the integral now
involves familiar-looking $P$ and $Q$ variables, the transformation we have
used will in general make $P$ and $Q$ complex, and the usual $P(t)$ and $Q(0)$
boundary conditions do not apply.  They are replaced by conditions on the
linear combinations $\alpha$ and $\bar\alpha$, reflecting the fact that
\(bfPQ) is still a transition amplitude between coherent states.  The $P,Q$
type of variables will be more convenient  for solving the equations of motion
of Section III, while $\alpha, \alphabar$ variables will still be better suited
to the initial and final states.  In the path integral formalism, there will be
no reason for us not to take the best of both worlds.

\subhead{\bf B.  Oscillator coupled to a heat bath}
Consider now a system consisting of a simple harmonic oscillator of natural
frequency $\Omega$, minimally coupled to a heat bath made up of environmental
oscillators.  This is a toy model, but it is very relevant to more significant
problems involving free quantum fields.  We assume the  Hamiltonian operators
$$\eqalign{
\hat H =& \hat H\-{SHO} + \hat H\-{ENV}\cr
\hat H\-{SHO} =& \12 (\hat P^2 + \Omega^2 \hat Q^2) = \Omega \,(\hat
a^\dagger\hat a + \12)\cr
\hat H\-{ENV} =& \12\int_0^\infty\!d\omega\, I(\omega)\Bigl(
({\hat p\-\omega\over I(\omega)} - g \hat Q)^2 +
\omega^2 \hat q\-\omega^2\Bigr)\;,}\eqno(Hsho)
$$
where $\hat a \equiv {1\over\sqrt{2\Omega}}(\Omega \hat Q + i \hat P)$, and $g$
is a coupling constant.  (We assume that $g^2 << \Omega$, and will use this
fact
extensively below.)  $I(\omega)$ is a dimensionless spectral density, whose
properties will be further discussed below.  Up to a rescaling of the
operators, \(Hsho) represents the choice of a minimal coupling similar to that
of standard electromagnetism.  This choice also has the
advantages of making $\hat H$ positive definite and of reducing the size of the
environmental (additive) re-normalization of $\Omega$.

We are interested in the behaviour of the observed oscillator alone, and so we
must integrate out of the problem all of the environmental degrees of freedom.
Assume that the initial density matrix for the complete system factorizes:
$$
{\bf \hat\rho}(0) = \hat\rho(0) \otimes \hat r(0)\;,\eqno(fact)
$$
where $\hat\rho(0)$ acts in the Hilbert space of the observed oscillator, and
$\hat r(0)$ is the initial density operator for the environment.
The reduced density matrix for the observed oscillator at time $t$ is
then given, in the coherent state basis, by the propagator equation
$$\eqalign{
\rho(\bar\alpha\-f, \alpha'\-f ;t)
 &\equiv \langle\bar\alpha\-f|\hat\rho(t)|\alpha'\-f\rangle\cr
 &=\int\!{d\alpha\-i d\alphabar'\-i d\xi d\bar\xi\over (2\pi i)^2}\,
      J(\bar\alpha\-f,\alpha'\-f,\alphabar'\-i,\alpha\-i;t)
      e^{-\bar\xi\alpha\-i} e^{-\alphabar'\-i\xi}
         \rho(\bar\xi, \xi ;0)\;.}
\eqno(propJ)
$$
This equation may be translated into the more familiar language of the energy
basis, by defining the propagator $K\-{klmn}$ in the basis of energy
eigenstates, such that
$$
J(\bar\alpha\-f, \alpha'\-f, \bar\alpha'\-i ,\alpha\-i ;t) \equiv \sum_{klmn}
K_{klmn}(t)
{\bar\alpha'^k\-i\alpha^l\-i\bar\alpha^m\-f\alpha'^n\-f\over\sqrt{k!l!m!n!}}\;,
\eqno(JK)
$$
which implies that
$$
\rho\-{mn}(t) = \sum_{k,l} \rho\-{kl}(0) K_{klmn}(t)\;.\eqno(propK)
$$

$J(\bar\alpha\-f,\alpha'\-f;\bar\alpha'\-i,\alpha\-i;t)$ may be
obtained explicitly from a path integral similar to \(bfPQ):
$$\eqalign{
J(\bar\alpha\-f,\alpha'\-f;\bar\alpha'\-i,\alpha\-i;t) =&
\int\!{\s D}Q{\s D}Q'{\s D}P{\s D}P'\,
\delta\bigl(\alpha(0) - \alpha\-i\bigr)
\delta\bigl(\bar\alpha(t) - \bar\alpha\-f\bigr)\cr
&\qquad\qquad\times\;\delta\bigl(\alpha'(t) - \alpha'\-f\bigr)
\delta\bigl(\bar\alpha'(0) - \bar\alpha'\-i\bigr)\, e^{i(A+B+V)}\;.}\eqno(JPI)
$$
In this expression, $A$ is the action term
$$
A \equiv\12\int_0^t\!ds\,P\dot{Q}-Q\dot{P} - P'\dot{Q}'+ Q'\dot{P}'
   - (P^2 - P'^2) - \Omega^2(Q^2-Q'^2) \; ;\eqno(Action)
$$
$B$ is the boundary term
$$
B\equiv {{\bar\alpha\-f\alpha(t) + \bar\alpha(0)\alpha\-i
        +\bar\alpha'\-i\alpha'(0) + \bar\alpha'(t)\alpha'\-f}\over2i}\;,
\eqno(Bdry)
$$
where the $\alpha$ variables are given in terms of the $Q$'s and $P$'s by
\(QP); and $V = V[Q,Q']$ is the influence phase.  (The quantity $e^{iV}$ is
known as  the {\it influence functional}.)

$V$ contains all the information about the heat bath that is relevant to the
observed oscillator.  It may be computed in a straightforward manner.  (This
has been done using the Lagrangian form of the environmental path integral: the
Lagrangian obtained from $H\-{ENV}$ is
$$
L\-{ENV} = \12\int_0^\infty\!I(\omega)d\omega\,\Bigl(\dot{q}\-\omega^2
+2gQ\dot{q}\-\omega -\omega^2q\-\omega^2\Bigr)\;,\eqno(Lenv)
$$
which is that of the velocity coupled model analysed in Reference
[\cite{CaldLegg}].) In the case where the initial state of the bath is thermal
at inverse temperature $\beta$, one finds that
$$\eqalign{
V[Q,&Q'] =\cr
 &{i g^2\over2}  \int_0^\infty\!d\omega\,\omega I(\omega)
 \int_0^t\!ds\int_0^s\!ds'\,
   (Q-Q')_s\Bigl((Q-Q')_{s'} \coth{\beta\omega\over2} \cos{\omega (s-s')}\cr
 &\qquad\qquad\qquad\qquad\qquad\qquad\qquad
 -i(Q+Q')_{s'} \sin{\omega (s-s')}\Bigr)\cr
 &\qquad\qquad
-{g^2\over2}\int_0^\infty\!d\omega\,I(\omega)\;\int_0^t\!ds\,(Q^2-Q'^2)_s\cr
\equiv&{i\over2} g^2\int_0^t\!ds\int_0^s\!ds'\, (Q-Q')\-s
\Bigl((Q-Q')_{s'}\nu (s-s')-i(Q+Q')_{s'}\eta (s-s')\Bigl)\cr
     &\qquad - {\mu^2\over2}\int_0^t\!ds\,(Q^2-Q'^2)\;.}\eqno(IF)
$$
We use the notation $(Q-Q')\-s = \bigl(Q(s)-Q'(s)\bigr)$. The last lines of
\(IF) implicitly define the quantity $\mu^2$ and the functions $\eta$ and
$\nu$.

The exponent $i(A+B+V)$ in \(JPI) is quadratic, and so \(JPI) is equal up to a
constant factor to the extremized value of the integrand.  Extremizing $A+B+V$
subject to the constraints imposed by the boundary conditions, we find the
first-order Euler-Lagrange equations for the four independent combinations
$Q\-\pm \equiv Q \pm Q'$, $P\-\pm \equiv P \pm P'$:
$$\eqalignno{
\dot{Q}\--(s) &= P\--(s)\; &(A)\cr
\dot{Q}\-+(s) &= P\-+(s)\; &(B)\cr
\dot{P}\--(s) &= -(\Omega^2+\mu^2)Q\--(s)
               -g^2\int_s^t\!ds'\,Q\--(s')\eta(s-s')\; &(C)\cr
\dot{P}\-+(s) &= -(\Omega^2+\mu^2)Q\-+(s)
               +g^2\int_0^s\!ds'\,Q\-+(s')\eta(s-s')\cr
           &\qquad\qquad +ig^2\int_0^t\!ds'\,Q\--(s')\nu(s-s')\;.&(D)}
$$
These equations together imply that $A+V$ vanishes; our problem therefore
reduces to one of endpoint values at the extremum:
$$\eqalign{
J(\bar\alpha\-f,\alpha'\-f,\bar\alpha'\-i,\alpha\-i;t) &= {1\over Z}
e^{i(A+B+V)}\Big\vert_0\cr
&={1\over Z}\exp{{\bar\alpha\-f\alpha(t)
+ \bar\alpha(0)\alpha\-i+\bar\alpha'\-i\alpha'(0) + \bar\alpha'(t)\alpha'\-f}
\over2}\Big\vert_0\;,}\eqno(Jsp)
$$
where $Z$ is a normalization constant.  Thus we only need to solve Eqs. \(A) --
\(D), subject to the boundary conditions
$$
\eqalign{
\12\bigl(\Omega(Q\-+ + Q\--) + i(P\-++P\--)\bigr)\Big\vert_{s=0} &=
\sqrt{2\Omega} \alpha\-i\cr
\12\bigl(\Omega(Q\-+ - Q\--) - i(P\-+-P\--)\bigr)\Big\vert_{s=0} &=
\sqrt{2\Omega} \bar\alpha'\-i\cr
\12\bigl(\Omega(Q\-+ - Q\--) + i(P\-+-P\--)\bigr)\Big\vert_{s=t} &=
\sqrt{2\Omega} \alpha'\-f\cr
\12\bigl(\Omega(Q\-+ + Q\--) - i(P\-++P\--)\bigr)\Big\vert_{s=t} &=
\sqrt{2\Omega} \bar\alpha\-f }\eqno(bc0)
$$
imposed by the delta functions in \(JPI), and find the remaining boundary
values that these solutions determine.

The easiest way to solve these coupled integro-differential equations is to
work in Fourier space.  The problem then reduces to finding the precise
location of a complex resonant frequency, as a pole in Fourier space.  To
begin, we differentiate \(A) and substitute it into \(C), obtaining a
decoupled, second order integro-differential equation for $Q\--$:
$$
\ddot{Q}\-- + (\Omega^2 + \mu^2)Q\-- = -g^2 \int_s^t\!ds'\,Q\--(s')
\eta(s-s')\;.\eqno(Qs)
$$
We then define the Fourier transform $Q\--(s) = {1\over2\pi i}
\int_{-\infty}^\infty\!d\omega\,
Q^-\-\omega e^{i\omega(t-s)}$, and use the definition $\eta(s-s')=
\int_0^\infty\!d\omega\,\omega I(\omega) \sin\omega(s-s')$ (implicit in
Eq. \(IF)) to obtain
$$
\int_{-\infty}^\infty\!d\omega\,e^{i\omega(t-s)} \left[Q^-\-\omega
\Bigl(\Omega^2 -
\omega^2\bigl(1+g^2\lambda(\omega^2)\bigr)\Bigr) - {g^2\over2}\omega
I(|\omega|)
P\!\int_{-\infty}^\infty\!d\omega'\, {Q^-\-{\omega'}\over \omega'-\omega}
\right]
= 0\;,\eqno(FQ)
$$
where we have combined the $\mu^2=\int_0^\infty\!d\omega\,I(\omega)$ term with
the contribution from the lower limit of the $s'$ integral to form the
frequency re-normalization term
$$
\lambda(\omega^2)\equiv P\!\int_0^\infty\!d\omega'\,
{I(\omega')\over \omega'^2-\omega^2}
\;.\eqno(lambda)
$$
The $P$ in front of the integral sign denotes the Cauchy principal value.  We
assume that $\lambda(\omega^2) << {1\over g^2}$.

The crux of the problem now appears to be the Hilbert transform in the
rightmost term of \(FQ).  The Hilbert transform has the property that
$$
P\!\int_{-\infty}^\infty\!{d\omega'\over \omega'-\omega}\
{1\over \omega'- (x + iy)} =
-i\pi\sgn{y} {1\over \omega - (x+iy)}\;.\eqno(Hilbert)
$$
With this in mind, we choose the ansatz
$$
Q\-\omega^- = {A\over \omega - (\bar\Omega + i\gamma)}\eqno(Qnu)
$$
for some $\bar\Omega$ and $\gamma >0$ to be determined, and $A$ a constant to
be fitted to boundary conditions.  Equation \(FQ) then becomes
$$
\int_{-\infty}^\infty\!d\omega\,{e^{i\omega(t-s)}\over \omega-(\bar\Omega +
i\gamma)}\Bigl( \omega^2(1+g^2\lambda) - \Omega^2 - i{\pi g^2\over2} \omega
I(|\omega|)\Bigr) = 0\;.\eqno(FQ2)
$$

We now invoke our assumption that $g^2\over\Omega$ is a small quantity, and
require further that the spectral
density $I(\omega)$ is always much smaller than ${\Omega\over g^2}$, and is
slowly varying near $\omega =\Omega$.  With these conditions, we can choose
$$\eqalign{
\gamma &= {\pi g^2\over 4}I(\Omega)\cr
\bar\Omega &= {\Omega\over\sqrt{1+g^2\lambda(\Omega^2)}}\;,}\eqno(wkgam)
$$
and take the $Q^-\-\omega$ chosen in \(Qnu) as the leading term of a Born
series
in powers of $g^2\over\Omega$\footnote{$^\dagger$}
   {This series may be constructed by
   replacing $I(\omega)$ with a sum over closely spaced delta functions.  (A
   continuous spectral density is realistic only as an approximation for such a
   discrete spectrum, anyway.)  Representing the delta functions as
   $\epsilon\over (\omega-\omega\-n)^2 + \epsilon^2$ then allows us to use
   \(Hilbert) repeatedly, order by order.}.
The first order correction to $Q\--(s)$ will then be
$$\eqalign{
{Ag^2\over4\pi i\Omega}&
\int_{-\infty}^\infty\!d\omega\,{e^{i\omega(t-s)}\over \omega-\bar\Omega -
i\gamma}\Bigl({i\pi\omega\over2}\bigl(I(\Omega) - I(|\omega|)\bigr)
+ \omega^2\bigl(\lambda(\omega^2)-\lambda(\Omega^2)\bigr)\Bigr)\cr
&\qquad\times\; \Bigl( {1\over\omega-\bar\Omega-i\gamma} -
{1\over \omega +\bar\Omega - i\gamma}\Bigr)\;.}\eqno(corr1)
$$
The poles at $\pm\bar\Omega+i\gamma$
contribute ``corrections'' that may be absorbed into the zeroth order solution.
For a wide range of possible spectral densities having no sharp
peaks near $\omega = \Omega$, the remaining correction will indeed be small.
In particular, we can have a gradual cut-off at high frequencies, which is
typical of realistic spectral densities.

We can substitute $\bar\Omega\to -\bar\Omega$ and the preceding remarks will
remain valid.  With our assumptions about $g^2$ and $I(\omega)$, then, we have
the  approximate solution to \(Qs)
$$
Q\--(s) \simeq Ae^{i\bar\Omega(t-s)}e^{\gamma(s-t)} +
\bar A e^{-i\bar\Omega(t-s)}e^{\gamma(s-t)}\;,\eqno(Qss)
$$
valid up to corrections of order ${\gamma\over\Omega}$.

This solution is anti-damped, growing exponentially with time.  Such behaviour
is usually considered unphysical, but the usual reasons for forbidding such
solutions do not apply in this case.  Despite the fact that $Q\--(s)$ is
treated similarly to a position variable, it is really a measure of the
departure from diagonality in the position basis of the oscillator's density
matrix, and it  is $Q\-+$ that represents  the oscillator's mean displacement.
Furthermore, \(Qs) is not really a  classical equation of motion, but rather a
tool for computing the evolution of a quantum mechanical system.  In classical
mechanics, where one's only boundary  conditions must apply at the initial
time, there is no natural way to get finite results from a ``runaway''
solution; but in quantum mechanics there is always a final time, with its own
boundary conditions, to which growing solutions may be made to conform.  And in
this case, the anti-damped behaviour of \(Qss) is the essential feature
responsible for establishing canonical thermal equilibrium, which has long been
known to  occur in the limit where ${g^2\over\Omega}\to 0$ but $\gamma
t\to\infty$[\cite{DvH}].

We now turn to Eq. \(B), differentiating it and applying \(D).  We obtain the
damped and driven oscillator equation
$$
\ddot{Q}\-+ + (\Omega^2 + \mu^2)Q\-+ - g^2\int_0^s\!ds'\,Q\-+(s')\eta(s-s') =
ig^2\int_0^t\!ds'\,Q\--(s')\nu(s-s')\;.\eqno(Q+s)
$$
The homogeneous part of \(Q+s) is simply the equation formed by substituting
$(t-s)\to s$ in \(Qs).  We therefore have the free (approximate) solutions
$$
Q^0\-+(s) \simeq Be^{-i\bar\Omega s}e^{-\gamma s}
+ \bar B e^{i\bar\Omega s}e^{-\gamma s}\;. \eqno(Q0ss)
$$

The source on the RHS of \(Q+s) is given by \(Qss) as
$$\eqalign{
&ig^2\int_0^t\!ds'\,Q\--(s')\nu(s-s') \simeq \cr
&{g^2\over2}\int_{-\infty}^\infty\!d\omega\,\omega\coth{\beta\omega\over2}
I(|\omega|) e^{i\omega s} \Bigl(A{e^{(i\bar\Omega-\gamma)t}-e^{-i\omega t}\over
\omega+\bar\Omega + i\gamma} + \bar A
{e^{(-i\bar\Omega-\gamma)t}-e^{-i\omega t}\over \omega -\bar\Omega +
i\gamma}\Bigr)\;.}\eqno(RHS)
$$
For the particular solution, we presume that $Q\-+(s)$ does behave as a damped
oscillator, and therefore consider the candidate
$$\eqalign{
Q^p\-+(s) =& {g^2\over2}\int_{-\infty}^\infty\!d\omega\,{\omega
I(|\omega|)\coth{\beta\omega\over2}e^{i\omega s}\over
(\omega+\bar\Omega-i\gamma)(\omega-\bar\Omega-i\gamma)}\cr
&\qquad\times \Bigl(A{e^{-i\omega t}-e^{(i\bar\Omega-\gamma)t}\over
\omega+\bar\Omega + i\gamma} + \bar A
{e^{-i\omega t}-e^{(-i\bar\Omega-\gamma)t}\over \omega -\bar\Omega +
i\gamma}\Bigr)\cr
\equiv&\int_{-\infty}^\infty\!d\omega\,e^{i\omega s} (AQ^A\-\omega + \bar A
Q^{\bar A}\-\omega)\;.}
\eqno(cand)
$$
We examine first the term proportional to $A$ in the LHS of \(Q+s):
$$\eqalign{
LHS(A) &= A\int_{-\infty}^\infty\!d\omega\,e^{i\omega s}\Bigl( Q^A\-\omega
\bigl(\Omega^2-\omega^2(1+g^2\lambda)\bigr) - {g^2\over2}\omega
I(|\omega|)P\!\!\int_{-\infty}^\infty\!d{\omega'}\,{Q^A\-{\omega'}
\over{\omega'}-\omega}\Bigr)\cr
&\simeq A\int_{-\infty}^\infty\!d\omega\,e^{i\omega s}\Bigl[Q^A\-\omega
\bigl(\Omega^2-\omega^2(1+g^2\lambda)\bigr) - {g^4\over4}\omega
I(|\omega|)\bar\Omega\coth{\beta\bar\Omega\over2} I(|\bar\Omega|)\cr
&\qquad\qquad\qquad\times
P\!\!\int_{-\infty}^\infty\!{d{\omega'}\over{\omega'}-\omega}
  {e^{-i{\omega'} t}-e^{(i\bar\Omega-\gamma)t}\over
({\omega'}+\bar\Omega +
i\gamma)({\omega'}+\bar\Omega-i\gamma)({\omega'}-\bar\Omega-i\gamma)}\Bigr]\cr
&=A\int_{-\infty}^\infty\!d\omega\,e^{i\omega s}\Bigl[Q^A\-\omega
\bigl(\Omega^2-\omega^2(1+g^2\lambda)\bigr) \cr
&\qquad\qquad\qquad
+ ig^2\gamma\omega {I(|\omega|)\bar\Omega\coth{\beta\bar\Omega\over2}
(e^{(i\bar\Omega-\gamma)t}-e^{-i\omega t})\over
(\omega+\bar\Omega + i\gamma)(\omega+\bar\Omega-i\gamma)
(\omega-\bar\Omega-i\gamma)}
\Bigr]\cr
&\simeq {Ag^2\over2}\int_{-\infty}^\infty\!d\omega\,e^{i\omega s}
{e^{(i\bar\Omega-\gamma)t}-e^{-i\omega t}\over\omega+\bar\Omega + i\gamma}
\;,}\eqno(LHA)
$$
neglecting terms of ${\s O}({g^2\over\Omega})$.  In \(LHA) we have twice used
the facts that ${1\over (\omega+\bar\Omega)^2 + \gamma^2}$ is sharply peaked
around $\omega=-\bar\Omega$, and that
$\omega\coth{\beta\omega\over2}I(|\omega|)$ is smooth in the same region.  We
have also invoked similar approximations to those used in deriving \(Qss).

Adding the similar term proportional to $\bar A$, and comparing the sum with
\(RHS), we can see that \(cand) does indeed satisfy \(Q+s) to leading order in
${g^2\over\bar\Omega}$.  Evaluating \(cand), once again using the  fact that
${1\over (\omega\pm\bar\Omega)^2 + \gamma^2}$ is sharply
concentrated, we find that
$$\eqalign{
\int_{-\infty}^\infty\!d\omega\,&Q^A\-\omega e^{i\omega s}\simeq\cr
  & -i\pi g^2\bar\Omega\coth{\beta\bar\Omega\over2}
I(\bar\Omega)
\Bigl({e^{(i\bar\Omega-\gamma)(t-s)}\over4i\gamma\bar\Omega}
 - {e^{(i\bar\Omega(t-s) -\gamma(t+s)}\over4i\gamma\bar\Omega} +
{e^{(i\bar\Omega-\gamma)(t+s)}\over4\bar\Omega^2}\Bigr)\;.}\eqno(three)
$$
But the decaying terms in \(three) are solutions to the homogeneous equation,
and may be absorbed into \(Q0ss).  Including the similar term proportional to
$\bar A$ (with a change of sign because $\coth{\beta\bar\Omega\over2}$ is odd),
we therefore write the approximate solution
$$
Q\-+(s) \simeq (Be^{-i\bar\Omega s} + \bar Be^{i\bar\Omega s})e^{-\gamma s}
-\coth{\beta\bar\Omega\over2}(Ae^{i\bar\Omega(t-s)} -
\bar A e^{-i\bar\Omega(t-s)})e^{-\gamma(t-s)}  \;.\eqno(Q+ss)
$$

Using \(Qss) and \(Q+ss) in \(A) and \(B), we find
$$\eqalign{
iP\--(s) \simeq& \Omega (Ae^{i\bar\Omega(t-s)}e^{\gamma(s-t)} -
\bar A e^{-i\bar\Omega(t-s)}e^{\gamma(s-t)})\cr
iP\-+(s) \simeq&
-\Omega\coth{\beta\bar\Omega\over2}\bigl(Ae^{i\bar\Omega(t-s)} +
 \bar A e^{-i\bar\Omega(t-s)}\bigr)e^{-\gamma(t-s)}\cr
& + (B e^{-i\bar\Omega s} -
\bar B e^{i\bar\Omega s})e^{-\gamma s}\;.}\eqno(Pss)
$$
We can now impose the boundary conditions of \(bc0):
$$\eqalign{
\sqrt{2\Omega} \alpha\-i &\simeq
\Omega\bigl((1-\coth{\beta\bar\Omega\over2})Ae^{(i\bar\Omega-\gamma)t} +
B\bigr)\cr
\sqrt{2\Omega} \alpha'\-f &\simeq \Omega\bigl(-(1+\coth{\beta\bar\Omega\over2})
A+ B e^{(-i\bar\Omega-\gamma)t} \bigr)\cr
\sqrt{2\Omega} \bar\alpha\-f &\simeq
\Omega\bigl((1+\coth{\beta\bar\Omega\over2})\bar A
+ \bar B e^{(i\bar\Omega-\gamma)t} \bigr)\cr
\sqrt{2\Omega} \bar\alpha'\-i &\simeq
\Omega\bigl((\coth{\beta\bar\Omega\over2}-1)
\bar Ae^{(-i\bar\Omega-\gamma)t} + \bar B\bigr)\;.}\eqno(bc)
$$
In the thermal limit where $e^{-\gamma t}\to 0$ and ${g^2\over\Omega}\to 0$,
these imply
$$\eqalign{
A&=-\sqrt{2\over\Omega}{\alpha'\-f\over \coth{\beta\bar\Omega\over2}+1}\cr
\bar A&=\sqrt{2\over\Omega}{\bar\alpha\-f\over
\coth{\beta\bar\Omega\over2}+1}\cr
B&=\sqrt{2\over\Omega}\alpha\-i\cr
\bar B &=\sqrt{2\over\Omega}\bar\alpha'\-i\;.}\eqno(ABCD)
$$

Using \(ABCD) and \(QP), we now obtain
$$\eqalign{
\alpha(t)&= e^{-\beta\bar\Omega}\alpha'\-f\cr
\bar\alpha'(t) &= e^{-\beta\bar\Omega}\bar\alpha\-f}
\qquad\qquad \eqalign{\alpha'(0) &= \alpha\-i\cr
\bar\alpha(0) &= \bar\alpha'\-i\;\;.}\eqno(alphas)
$$
Applying these results to \(Jsp), we find the propagator
$$
J = {1\over Z} \exp{ \bigl(\bar\alpha'\-i\alpha\-i +
e^{-\beta\bar\Omega}\bar\alpha\-f\alpha'\-f\bigr)}\;,\eqno(Jss)
$$
where $Z$ will turn out to be the partition function of the canonical
ensemble.  Using \(propJ), we find that the final state of the oscillator has
the density matrix
$$
\rho(\bar\alpha,\alpha) = {1\over Z}\exp
\bigl(e^{-\beta\bar\Omega}\bar\alpha\alpha\bigr)\;,
$$
which is the canonical ensemble in coherent state variables,
regardless of the initial state.  We may translate this result into the
standard energy representation, using \(JK) to discern
$$
K\-{klmn}(t) = (1-e^{-\beta\bar\Omega}) \delta\-{kl}
e^{-m\beta\bar\Omega}\delta\-{mn}
+ {\s O}({g^2\over\Omega}) + {\s O}(e^{-\gamma t})\;,
$$
which implies that
$$
\rho_{mn}(t) \simeq (1-e^{-\beta\Omega})\delta_{mn}
e^{-m\beta\Omega}\;.\eqno(rhomn)
$$
(Although $\bar\Omega = \Omega + {\s O}({g^2\over\Omega})$, one might be
concerned that for low temperatures, where $\beta$ becomes arbitrarily large,
$\beta g^2\lambda(\Omega^2)$ would not be small, so that it would be wrong to
write $\Omega$ instead of $\bar\Omega$ in \(rhomn).  But because $g^2 <<
\Omega$, regardless of temperature, $\beta g^2$ can only be non-negligible when
$\beta\Omega$ is so large that $e^{-\beta\bar\Omega}\simeq  e^{-\beta\Omega}
\simeq 0$.  Hence the distinction between $\Omega$ and $\bar\Omega$ is not
discernible in the final state, within the limits of our approximations.)

We therefore conclude that a harmonic oscillator, weakly damped by coupling to
an environment with slowly varying spectral density, is inexorably driven to
thermal equilibrium on the thermal time scale $t \sim {1\over\gamma}$.  This
extends the result of Reference [\cite{Grabert}], where the path integral in
position variables was used to obtain thermalization for a weakly  coupled
environment with ohmic spectral density.  (In our velocity coupled model, this
would be equivalent to constant $I(\omega)$.  Reference [\cite{Grabert}]
notes that more general dissipative environments can induce thermalization, but
does not treat them explicitly.)  We conclude this section by emphasizing that,
in the case of the single oscillator coupled to a bath, it is the long time
scale that makes the thermal effects of the bath dominate the system, and it is
weak coupling and  slowly varying spectral density that effectively prevent the
environment from communicating to the system any information other than its
temperature, and so guarantee the canonical thermal behaviour.

\head{\bf III.  Black body radiation}

\subhead{\bf A. Setting up the model}

We now present our main result.  When a massless field is weakly coupled to an
environment which is confined to a finite region of space, modes with
wavelengths long compared to the coupling tend to be reflected at  the
boundaries of this region, but shorter wavelength quanta may experience
negligible reflection.  When the thickness of the region is large compared to
the coupling, transmission through it is also negligible, and the quanta are
absorbed --- and emitted as thermal radiation.  After the initial time $s=0$,
when the field and its environment are uncorrelated, this radiation propagates
outwards from the  absorbing region; but behind a wavefront region whose width
is on the order of the inverse square of the  coupling,  it is
time-independent\footnote{$^\dagger$}{Cooling of the emitting medium is not
considered in this section.}.   While a coloured environment (\ie one with
slowly  varying spectral density $I(\omega)$) may be analysed by
straightforwardly  extending the approach of Section II, for simplicity in this
Section we consider an ohmic spectral density ($I(\omega) = 1$).  The
environment we discuss thus represents a black body, absorbing all frequencies
(above the reflection regime) indiscriminately.

Also for simplicity, we consider a scalar field in $1+1$ dimensions; the
extrapolation to a vector field in $3+1$ dimensions is straightforward
enough, but will only be touched on in this paper.
We let our spatial co-ordinate $x \in [-\Lambda, \Lambda]$, and impose periodic
boundary conditions on the field $\phi$.  (We will eventually let $\Lambda \to
\infty$, of course, but this infrared regulator will turn out to be
convenient.)   The black body is placed at the origin, filling the region
$-L < x < L$ with a uniform medium consisting of independent ohmic heat baths
at every point, minimally coupled to the field\footnote{$^\ddagger$}
{Attaching an independent bath
to every point is an approximation, valid for sufficiently long wavelengths, as
discussed in Section IV.  Similar models, at zero temperature, have
been discussed in References [\cite{atten}] and [\cite{Candelas}].}.
The Hamiltonian is
$$\eqalign{
\hat H =& \12\lint dx\, \bigl(\hat\Pi^2(x) + \partial\-x\hat\phi^2(x)\bigr)\cr
&+ \int_{-L}^L\!dx\,\int_0^\infty\!d\omega\,
\Bigl(\bigl(\hat{p}\-\omega(x)-g\hat{\phi}(x)\bigr)^2
+ \omega^2 \hat{q}^2\-\omega(x)\Bigr)\;\;.}\eqno(Ham)
$$
Once again, we assume that $g^2$ is very small compared to an infrared cut-off
frequency $\Omega$.

Define creation and annihilation operators by
$$
\hat a\-n =  \lint{dx\over2\sqrt{|n|\pi}}\,e^{-i{n\pi\over\Lambda}x}
\bigl(\big\vert{n\pi\over\Lambda}\big\vert\hat\phi(x) + i\hat\Pi(x)\bigr)\;,
$$
so that $[\hat a\-n,\hat a^\dagger\-m] = \delta\-{mn}$, and then introduce the
Bargmann-Fock variables $\alpha\-n(s)$ and  $\alphabar\-n(s)$. (For the mode
$n=0$, we are proceeding as though the field had an infinitesimal IR regulating
mass $\epsilon<<{\pi\over\Lambda}$, so that
$$
{n\pi\over\Lambda} \to \left\{
\eqalign{{n\pi\over\Lambda}\;,\;n\neq &0\cr  \epsilon\;,\; n=&0\;\;;}\right.
$$
but we will keep this replacement implicit for brevity.) Introduce as well the
primed variables $\alpha'\-n(s), \alphabar'\-n(s)$ needed to express a mixed
state of the field.

In this section we once again use a path integral with an influence functional
to derive a propagator, similar to that in \(propJ), but for the density
operator of the scalar field:
$$\eqalign{
{\s \rho}[\alphabar^f\-m, \alpha'^f\-n; t] &\equiv
\langle \alphabar^f\-m|\;\hat{\s \rho}(t)|\alpha'^f\-n\rangle\cr
&= \int\!\prod_{k,l=-\infty}^\infty
{d\alpha^i\-k d\bar\xi\-k\over2\pi i}{d\alphabar'^i\-l d\xi\-l\over2\pi i}\,
e^{-\sum_k(\bar\xi\-k\alpha^i\-k + \alphabar'^i\-k\xi\-k)}\cr
&\qquad\qquad\times\,
{\s J}[\alpha^i\-k, \alphabar'^i\-l, \alphabar^f\-m,\alpha'^f\-n;t]
\,{\s \rho}[\bar\xi\-k, \xi\-l; 0]\;,}\eqno(sJ)
$$
where $|\alpha'^f\-n\rangle$ is shorthand for the infinite tensor product state
$\prod_{n=-\infty}^\infty\!|\alpha'^f\-n\rangle$, and likewise for the bra.

The propagator ${\s J}$ may be computed from a path integral somewhat similar
to \(JPI):
$$
{\s J}[\alpha^i\-k, \alphabar'^i\-l, \alphabar^f\-m,\alpha'^f\-n;t]
=\int\!\D\phi \D\phi' \D\Pi \D\Pi' \, e^{i({\s A + B + V})}\;,\eqno(sJpi)
$$
where the field variables $\phi(x,s)$ and $\Pi(x,s)$ are defined from the
Bargmann-Fock variables according to
$$\eqalign{
\alpha\-m(s) =&  \lint{dx\over2\b{m}}\, e^{-im{\pi x\over\Lambda}}\bigl(
|{m\pi\over\Lambda}|\phi(x,s) + i\Pi(x,s)\bigr)\cr
\alphabar\-m(s) =& \lint{dx\over2\b{m}}\, e^{im{\pi x\over\Lambda}}\bigl(
|{m\pi\over\Lambda}|\phi(x,s) - i\Pi(x,s)\bigr)\;.}\eqno(phiPi)
$$
The primed version of \(phiPi) defines $\phi'$ and $\Pi'$.
The boundary conditions implicit in \(sJpi) are as follows:
$$\eqalign{
\alpha\-m(0)= \lint{dx\over2\b{m}}\, e^{-im{\pi x\over\Lambda}}\bigl(
|{m\pi\over\Lambda}|\phi(x,0) + i\Pi(x,0)\bigr)&=\alpha^i\-m\;\;;\cr
\alphabar\-m(t)=\lint{dx\over2\b{m}}\, e^{im{\pi x\over\Lambda}}\bigl(
|{m\pi\over\Lambda}|\phi(x,t) - i\Pi(x,t)\bigr)&=\alphabar^f\-m\;\;;\cr
\alphabar'\-m(0)= \lint{dx\over2\b{m}}\, e^{im{\pi x\over\Lambda}}\bigl(
|{m\pi\over\Lambda}|\phi'(x,0) - i\Pi'(x,0)\bigr)&=\alphabar'^i\-m\;\;;\cr
\alpha'\-m(t)=\lint{dx\over2\b{m}}\, e^{-im{\pi x\over\Lambda}}\bigl(
|{m\pi\over\Lambda}|\phi'(x,t) + i\Pi'(x,t)\bigr)&=\alpha'^f\-m\;\;.}
\eqno(bcphiPi)
$$

The boundary term ${\s B}$ in \(sJpi) is most briefly written as
$$
{\s B} = {1\over2i}\sum_{m=-\infty}^\infty\bigl(\alphabar^f\-m\alpha\-m(t) +
\alphabar\-n(0)\alpha^i\-n + \alphabar'^i\-n\alpha'\-n(0)
+\alphabar'\-n(t)\alpha'^f\-n\bigr)\;,\eqno(sB)
$$
where the boundary values of the $\alpha$'s may be found from the boundary
values of the $\phi$'s and $\Pi$'s using \(phiPi).
As in Section II, the action term in the path integral is of the usual
$p\dot{q}-H$ form (up to a boundary term), minus a like term in primed
variables:
$$\eqalign{
{\s A} =& \12\int_0^t\!ds\!\lint dx\,\bigl(\Pi\dot\phi -
\phi\dot\Pi -\Pi'\dot\phi'+ \phi'\dot\Pi' - \Pi^2 + \Pi'^2 - \partial\-x\phi^2
+
\partial\-x\phi'^2\bigr)\cr
=&\12\int_0^t\!ds\!\lint dx\,\Bigl({\Pi\-+\dot{\phi}\-- + \Pi\--\dot{\phi}\-+ -
\phi\-+\dot{\Pi}\-- - \phi\--\dot{\Pi}\-+ \over 2} - \Pi\-+\Pi\-- -
\partial\-x\!\phi\-+\partial\-x\!\phi\--\Bigr)\;,}\eqno(sA)
$$
where we introduce the linear combinations $\phi\-\pm = \phi \pm \phi'$ and
$\Pi\-\pm = \Pi \pm \Pi'$.

Because the heat baths coupled to the field at each point are independent,
the influence phase ${\s V}$ is simply an integral of influence phases similar
to \(IF) (with $I(\omega)$ set equal to one and some simplification via
integration by parts, as in Reference [\cite{CaldLegg}]):
$$\eqalign{
{\s V} =& {ig^2\over8}\int_0^t\!ds\!\int_0^t\!ds'\int_{-L}^L\!dx\,\phi\--(x,s)
\phi\--(x,s')\int_{-\infty}^\infty\!d\omega\,\omega\coth{\beta\omega\over2}
e^{i\omega(s-s')}\cr
&\qquad
- {\pi g^2\over4}\int_0^t\!ds\int_{-L}^L\!dx\,\dot{\phi}\-+(x,s)\phi\--(x,s)\cr
&\qquad -{\pi g^2\over4}\int_{-L}^L\!dx\,\phi\-+(x,0)\phi\--(x,0)\;.}
\eqno(sIF)
$$
Since we will once again be interested in the case where $g^2$ is small, we
will neglect the last term in \(sIF), on the grounds that it will produce only
negligible perturbations on the solutions to the field equations.
The black body is here assumed to have a uniform initial temperature ${1\over
k\-B \beta}$, and the limits on the integral over $\omega$ are meant to be
taken to infinity only after all other integrations have been performed.

The exponent of the integrand in \(sJpi) is thus quadratic, and the propagator
${\s J}$ is proportional to the value of the integrand when this exponent is
extremized, subject to the boundary conditions.  Once again, ${\s A} + {\s V}$
vanishes on shell, and the propagator is given by
$$
{\s J}[\alpha^i\-k, \alphabar'^i\-l, \alphabar^f\-m,\alpha'^f\-n;t]
= {1\over Z} e^{i{\s B}\big\vert\-0}\;.\eqno(JBB)
$$
The Euler-Lagrange equations for this extremum are
$$\eqalignno{
\dot{\phi}\--(x,s) &= \Pi\--(x,s) &(dphm)\cr
\dot{\phi}\-+(x,s) &= \Pi\-+(x,s) &(dphp)\cr
\dot{\Pi}\--(x,s)  &= {d^2\ \over dx^2}\phi\--(x,s)
         + {\pi g^2\over2}\theta(L-|x|)\dot{\phi}\--(x,s) &(dpim)\cr
\dot{\Pi}\-+(x,s) &= {d^2\ \over dx^2}\phi\-+(x,s)
               -{\pi g^2\over2}\theta(L-|x|)\dot{\phi}\-+(x,s)\cr
               &\qquad +
       {ig^2\over2}\int_0^t\!ds'\,\phi\--(x,s')\int_{-\infty}^\infty\!d\omega\,
           \omega\coth{\beta\omega\over2} e^{i\omega(s-s')}\;\;.&(dpip)}
$$
Differentiating \(dphm) and \(dphp), and employing the results in \(dpim) and
\(dpip), we obtain the second order equations
$$\eqalignno{
\ddot{\phi}\-- - {d^2\ \over dx^2}\phi\-- - {\pi g^2\over2}\theta(L-|x|)
\dot{\phi}\-- &= 0 &(ddm)\cr
\ddot{\phi}\-+ - {d^2\ \over dx^2}\phi\-+ + {\pi g^2\over2}\theta(L-|x|)
\dot{\phi}\-+ &= \cr
{ig^2\over2}\int_0^t\!ds'\,\phi\--&(x,s')\int_{-\infty}^\infty\!
           d\omega\,\omega\coth{\beta\omega\over2}
e^{i\omega(s-s')}\;\;.&(ddp)}
$$

\subhead{\bf B. Solving the equations of motion}

We will find solutions to \(ddm) and \(ddp) that are complex exponentials in
time and space.  We begin with \(ddm), and consider a solution of the form
$$
\phi\--(x,s) = e^{-i k s} \left\{ \eqalign{L\-+ e^{i k x}
+ L\-- e^{-i k x}\qquad & x\in [-\Lambda, -L]\cr
M\-+ e^{i\kappa x} + M\-- e^{-i\kappa x} \qquad & x\in [-L,L]\cr
R\-+ e^{i k x} + R\-- e^{-i k x}\qquad & x\in [L,\Lambda] \;\;,}\right.
\eqno(LMR)
$$
where $ k$ is a complex frequency to be determined.
{}From \(ddm), we see that $\kappa$ must satisfy
$$
\kappa^2 =  k^2 - i{\pi g^2\over2} k = ( k - i{\pi g^2\over4})^2 \times
\bigl( 1 + {\s O}({g^4\over k^2})\bigr)\;.\eqno(Kappa)
$$

To be a saddlepoint of the path integral, the solution prescribed by \(LMR)
must be $C^1$ everywhere.  We must therefore match both $\phi\--$ and ${d\
\over dx}\phi\--$ as we approach $x=\pm L$ from either side.  It is easy to
see from the discontinuity in the wavenumber implied by \(Kappa) that there
will be reflection at the boundary, with reflection co-efficient of order
${g^2\over k}$.  This implies that there is really no such thing as a
perfect absorber, since below some threshold frequency there will
always be significant reflection.  Nevertheless, if we restrict our attention
to field modes with
frequencies above an infrared cut-off $\Omega$, such that $g^2<<\Omega$, then
we can neglect reflection, and consider our black body to absorb
perfectly at all frequencies of interest.  Satisfying the periodic boundary
conditions at $x=\pm\Lambda$ forces $ k$ to assume a discrete set of values.
Imposing these constraints on $L\-\pm$, $M\-\pm$, and $R\-\pm$, we find the
most convenient parametrization of the general solution, above the IR cut-off,
to be given approximately by
$$
\phi\--(x,s) = \sum_{{|n|\pi\over\Lambda} > \Omega}
e^{i{n\pi\over\Lambda}x}\bigl( B\-n f\-n(x)
e^{-i|{n\pi\over\Lambda}|s} + \bar B\-n f\-n(-x) e^{i|{n\pi\over\Lambda}|s}
\bigr) e^{\gamma s} + {\s O}({g^2\over\Omega})\;,\eqno(phin)
$$
where
$$\eqalign{
f\-n(x) &= f(x \sgn{n})\cr
f(x) &= \left\{\eqalign{
e^{-(\Gamma + \gamma)L} e^{-\gamma x} \qquad & -\Lambda <x<-L\cr
e^{\Gamma x} \qquad & -L<x<L\cr
e^{(\Gamma + \gamma)L} e^{-\gamma x} \qquad & L<x<\Lambda}\right.\cr
&= {1\over f(-x)}\;.}\eqno(f)
$$
The attenuation co-efficients $\gamma \equiv {\pi g^2\over4}{L\over\Lambda}$
and $\Gamma \equiv {\pi g^2\over4}{\Lambda-L\over\Lambda}$ are defined so that
$f$ is indeed periodic.

The general solution to the homogeneous part of \(ddp) is of the same form as
$\phi\--(x,-s)$.
The driven solutions, using the source $\phi\--(x,s')$ as given by \(phin),
may be found by the usual Fourier technique.  Once again discarding from this
particular solution terms which are actually free solutions, we obtain the
general approximate solution
$$\eqalign{
\phi\-+(x,s) &\simeq \sum_{|{n\pi\over\Lambda}|>\Omega}
e^{i{n\pi\over\Lambda}x}\bigl( B\-n f\-n(-x)e^{-i|{n\pi\over\Lambda}|s}
+ \bar B\-n f\-n(x) e^{i|{n\pi\over\Lambda}|s}\bigr) e^{-\gamma s}\cr
&\; - \sum_{|{n\pi\over\Lambda}|>\Omega} e^{i{n\pi\over\Lambda}x}
\coth{\beta |n|\pi\over2\Lambda}\bigl( A\-n f\-n(x)
e^{-i|{n\pi\over\Lambda}|s}
- \bar A\-n f\-n(-x) e^{i|{n\pi\over\Lambda}|s}\bigr)
e^{\gamma s}\;,}\eqno(phisol)
$$
where the ``$\simeq$'' means that we have neglected corrections of order
${g^2\over\Omega}$,  as well as all modes with frequency less than
$\Omega$.

Using \(phisol) in \(dphm) and \(dphp), and dropping terms of order
$\gamma\over\Omega$, we find also
$$\eqalign{
i\Pi\--(x,s) &\simeq \sum_{|{n\pi\over\Lambda}|>\Omega}
\big\vert {n\pi\over\Lambda}\big\vert
e^{i{n\pi\over\Lambda}x}\bigl( A\-n f\-n(x)e^{-i|{n\pi\over\Lambda}|s}
- \bar A\-n f\-n(-x) e^{i|{n\pi\over\Lambda}|s}\bigr) e^{\gamma s}\cr
i\Pi\-+(x,s) &\simeq \sum_{|{n\pi\over\Lambda}|>\Omega}
\big\vert {n\pi\over\Lambda}\big\vert
e^{i{n\pi\over\Lambda}x}\bigl( B\-n f\-n(-x)e^{-i|{n\pi\over\Lambda}|s}
- \bar B\-n f\-n(x) e^{i|{n\pi\over\Lambda}|s}\bigr) e^{-\gamma s}\cr
 - \sum_{|{n\pi\over\Lambda}|>\Omega} &\big\vert {n\pi\over\Lambda}\big\vert
e^{i{n\pi\over\Lambda}x}
\coth{\beta |n|\pi\over2\Lambda}\bigl( A\-n f\-n(x)
e^{-i|{n\pi\over\Lambda}|s}
+ \bar A\-n f\-n(-x) e^{i|{n\pi\over\Lambda}|s}\bigr) e^{\gamma s}\;.}
\eqno(Pisol)
$$
With regard to the neglect of modes below the IR cut-off $\Omega$, note that
these modes still have solutions of the form \(LMR), though with long
wavelengths and non-negligible reflection.  When we return to the Bargmann-Fock
variables using \(phiPi), therefore, their contribution to the Fourier
modes involved in the $\alpha\-m$ will be negligible, for
$|{m\pi\over\Lambda}|>\Omega$.  Hence the infrared modes are effectively
decoupled from the absorbed modes, and as long as we restrict our final results
to $\alpha\-m$ above the cut-off, we can use \(phisol) and \(Pisol)
without concern for the absence or presence of the  infrared contributions.

This last argument hinges on the fact that a complex exponential
$e^{(ik\pm\eta) x}$, with $|\eta| << |k|$ so that the growth or decay of the
function is only significant over very many wavelengths, is primarily composed
of Fourier modes $e^{ik'x}$ with $k'$ close to $k$.  This statement may be made
more explicit by considering the matrix
$$\eqalign{
d\-{mn} &\equiv \lint {dx\over2\Lambda} e^{i(n-m){\pi x\over\Lambda}}
f\-n(x)\cr
&= {e^{\pm\Gamma L}e^{i{(n-m)\pi\over\Lambda}L}
- e^{\mp\Gamma L}e^{-i{(n-m)\pi\over\Lambda}L}\over 2i\Lambda}
\left({1\over {(n-m)\pi\over\Lambda} \pm i\gamma}
- {1\over {(n-m)\pi\over\Lambda} \mp i\Gamma}\right)\;,}\eqno(dmn)
$$
where we take the upper (lower) of the $\pm$ signs for $n>0$ ($n<0$).
This matrix is {\it quasi-diagonal} --- predominantly concentrated within a
distance $|m-n| \sim
\Gamma\Lambda$ of the diagonal.  If we let ${g^2\over\Omega} \to 0$ while
keeping $\Gamma L$ finite, it approaches the Kronecker delta.  In fact, the
case where ${g^2\over\Omega}e^{\Gamma L}$ is not small can also be
encompassed, since the large factor $e^{\Gamma L}$ will cancel in our final
results, leaving the contributions from off-diagonal elements of $d\-{mn}$
still suppressed by $g^2\over\Omega$.  This will also ensure that reflection
terms ignored in \(phin) remain negligible, despite the fact that some will
have amplitudes of order $e^{\Gamma L}$.

The quasi-diagonality of $d\-{mn}$ will be of crucial importance in the
remainder of this paper.  We will use it to construct an approximation
whereby modes of all frequencies above the IR cut-off are slowly modulated, so
that they effectively vanish in large regions of space, but retain well-defined
wavelengths in the regions where they do not.  It is in this manner that we
will identify thermal radiation of all frequencies propagating outwards from
our
model black body.  The ``no reflection'' condition that ${g^2\over\Omega}<<1$
thus turns out to have additional simplifying consequences far beyond making
reflection negligible.  As in Section II, it is weak coupling and slowly
varying spectral density that lead to
canonical thermal behaviour; the non-reflectivity of a black body is simply a
sign that the weak coupling limit applies.

We can now re-combine $\phi\-\pm$ and $\Pi\-\pm$ into the Bargmann-Fock
$\alpha$ variables, having effectively used the other variables to de-couple
the Bargmann-Fock equations of motion. Combining \(phisol) and \(Pisol) with
the  definition \(phiPi), assuming  $|{m\pi\over\Lambda}| > \Omega$, and then
using $|n| d\-{mn}  \simeq |m| d\-{mn}$,  we deduce the following approximate
solutions for the Bargmann-Fock variables:
$$\eqalign{
\alpha\-m(s)&\simeq\b{m}
\sum_{n=-\infty}^\infty e^{-i|{n\pi\over\Lambda}|s}
\left( - d\-{mn}(\coth{\beta |n|\pi\over2\Lambda} - 1) e^{\gamma s} A\-n
+ d^*\-{mn} e^{-\gamma s} B\-n \right)\cr
\alpha'\-m(s)&\simeq\b{m}
\sum_{n=-\infty}^\infty e^{-i|{n\pi\over\Lambda}|s}
\left( - d\-{mn}(\coth{\beta |n|\pi\over2\Lambda} + 1) e^{\gamma s} A\-n
+ d^*\-{mn} e^{-\gamma s} B\-n \right)\cr
\alphabar\-m(s)&\simeq\b{m}
\sum_{n=-\infty}^\infty e^{i|{n\pi\over\Lambda}|s}
\left( d^*\-{mn}(\coth{\beta |n|\pi\over2\Lambda} + 1) e^{\gamma s} \bar A\-n
+ d\-{mn} e^{-\gamma s} \bar B\-n \right)\cr
\alphabar'\-m(s)&\simeq\b{m}
\sum_{n=-\infty}^\infty e^{i|{n\pi\over\Lambda}|s}
\left(  d^*\-{mn}(\coth{\beta |n|\pi\over2\Lambda} - 1) e^{\gamma s} \bar A\-n
+ d\-{mn} e^{-\gamma s} \bar B\-n \right)\;\;.}\eqno(alpham)
$$
Once again, the dotted equality sign indicates that these equations are valid
up to corrections of order $g^2\over\Omega$.

\subhead{\bf C. Boundary conditions}

We must now constrain $A\-n, \bar A\-n, B\-n, \bar B\-n$ to meet the boundary
conditions of \(bcphiPi).  We begin with $A\-n$ and $B\-n$.  The constraints on
$\alpha\-m(0)$ and $\alpha'\-m(t)$ may be decoupled by defining
$$\eqalign{
A\-n =& {1\over \coth{\beta |n|\pi\over2\Lambda} + 1}\sum_{k=-\infty}^\infty
\left( e^{i|{n\pi\over\Lambda}|t}d^{-1}\-{nk} e^{-\gamma t}C\-k -
d^{-1}\-{nk}D\-k\right)\cr
B\-n =& \sum_{k=-\infty}^\infty \left(e^{i|{n\pi\over\Lambda}|t} d^{*-1}\-{nk}
e^{\gamma t} C\-k - e^{-\beta |{k\pi\over\Lambda}|} d^{*-1}\-{nk}
D\-k\right)\;.}\eqno(CD)
$$
Using the quasi-diagonality of $d\-{mn}$, we can see that its inverse is given
approximately by
$$
d^{-1}\-{nm}\bigl(1+{\s O}({g^2\over\Omega})\bigr) =
\lint {dx\over2\Lambda} e^{i(n-m){\pi x\over\Lambda}}f\-n(x)\;,\eqno(dinv)
$$
since the matrix product $\sum_n d\-{mn} d^{-1}\-{nk}$ will, for
$|{n\pi\over\Lambda}| > \Omega$, be dominated by contributions where $m$, $n$,
and $k$ all have the same sign.  This allows us to substitute $f\-n(x) \to
f\-m(x)$ or $f\-k(x)$ with negligible error.

The inverse of $d\-{mn}$ is thus nearly diagonal as well, and so
$d^{*-1}\-{nk}e^{-\beta{|k|\pi\over\Lambda}} \simeq
d^{*-1}\-{nk}e^{-\beta{|n|\pi\over\Lambda}}$ (since if $\beta$ is large enough
for $\Gamma\beta$ to be non-negligible, then $e^{-\beta\Omega}\to 0$ and the
whole term is negligible).  This allows us to write the decoupled conditions
$$\eqalign{
\alpha^i\-m &\simeq \b{m} \sum_{k,n}\bigr(
d^*\-{mn} e^{i{|n|\pi\over\Lambda}t} d^{*-1}\-{nk} e^{\gamma t}
- d\-{mn} e^{-\beta{|n|\pi\over\Lambda}}
e^{i{|n|\pi\over\Lambda}t} d^{-1}\-{nk} e^{-\gamma t}
\bigr) C\-k\cr
\alpha'^f\-m &\simeq \b{m} \sum_{k,n}\bigr(
d\-{mn} e^{-i{|n|\pi\over\Lambda}t} d^{-1}\-{nk} e^{\gamma t}
- d^*\-{mn} e^{-\beta{|n|\pi\over\Lambda}}
e^{-i{|n|\pi\over\Lambda}t} d^{*-1}\-{nk} e^{-\gamma t}
\bigr) D\-k\;.}\eqno(alphaif0)
$$

We can expand \(alphaif0) into
$$\eqalign{
\alpha^i\-m&\simeq \b{m} \sum_{k,n} C\-k \lint {dx dy\over (2\Lambda)^2}\,
e^{i(mx-ky){\pi\over\Lambda}} e^{-{i\pi n\over\Lambda}(x-y-t\sgn{n})}
\cr
&\qquad\qquad \times \bigl( f\-n(x)f\-n(-y)e^{\gamma t} - f\-n(-x)f\-n(y)
e^{-\beta{|n|\pi\over\Lambda}}e^{-\gamma t}\bigr)\cr
&\simeq \b{m} \sum_{k} C\-k \lint {dx\over2\Lambda}
e^{i(m-k){\pi x\over\Lambda}}
e^{i|k|{\pi t\over\Lambda}}\bigl(G\-m(x,t) - {e^{-\beta{|m|\pi\over\Lambda}}
\over G\-m(x,t)}\bigr)\cr
\alpha'^f\-m&\simeq \b{m} \sum_{k} D\-k \lint {dx\over2\Lambda}
e^{-i(m-k){\pi x\over\Lambda}}
e^{-i|k|{\pi t\over\Lambda}}\bigl(G\-m(x,t) -
{e^{-\beta{|m|\pi\over\Lambda}}\over G\-m(x,t)}\bigr)\;,}
\eqno(alphaif)
$$
where $G\-m(x,t) \equiv e^{\gamma t}
f\-m(x) f\-m(-x + t\,\sgn{m})$ (continuing
$f$ periodically beyond $[-\Lambda, \Lambda]$).
These last two lines are obtained via another quasi-diagonal approximation:
because the dominant contributions to
\(alphaif0) come from the terms where $m$, $n$, and $k$ are all within
$\Gamma\Lambda$ of each other, and therefore (for $|{m\pi\over\Lambda}| >
\Omega$) all have the same signs,
we can switch $f\-n \to f\-m$, and $|n| \to n \sgn{m}$ or
$n\sgn{k}$, and incur errors of order $g^2\over\Omega$ at worst.  This type of
approximation will be used
repeatedly hereafter, drastically simplifying several matrix inversions.
In effect, the only reason not to treat $d\-{mn}$ as a delta function is that,
for large enough $t$, $e^{i{n\pi\over\Lambda}t}$ can vary significantly over
its otherwise negligible width.
For sufficiently small $g^2\over\Omega$, this quasi-diagonal approximation
is accurate and natural.  At the quantum
level, the quasi-diagonal approximation is the ideal black body approximation.

Applying this approximation again allows us to invert \(alphaif), obtaining
$$\eqalign{
C\-k &\simeq  \sum_{m} {\alpha^i\-m\over\b{m}}
\lint {dx\over2\Lambda} {e^{i(k-m){\pi x\over\Lambda}}
e^{-i|k|{\pi t\over\Lambda}}\over
G\-m(x,t) - {e^{-\beta{|m|\pi\over\Lambda}}\over G\-m(x,t)}}\cr
D\-k &\simeq \sum_{m} {\alpha'^f\-m\over\b{m}}
\lint {dx\over2\Lambda} {e^{-i(k-m){\pi x\over\Lambda}}
e^{i|k|{\pi t\over\Lambda}}\over
G\-m(x,t) - {e^{-\beta{|m|\pi\over\Lambda}}\over G\-m(x,t)}}\;.}\eqno(DCinv)
$$
Note that the function $G\-m(x,t) = G(x\,\sgn{m},t)$, where for
$t>2L$
$$
G(x,t) = f(x) f(-x+t) e^{\gamma t} = \left\{\eqalign{
1\;,\qquad & x\in [-\Lambda, -L]\cr
e^{{\pi\over4}g^2(L+x)}\;,\qquad & x\in [-L, L]\cr
e^{{\pi\over2}g^2L}\;,\qquad & x\in [L, t-L]\cr
e^{{\pi\over4}g^2(L+t-x)}\;,\qquad & x\in [t-L, t+L]\cr
1\;,\qquad & x\in [t+L, \Lambda]}\right.\eqno(Gxt)
$$
For $t<2L$, we have instead
$$
G(x,t) = f(x) f(-x+t) e^{\gamma t} = \left\{\eqalign{
1\;,\qquad & x\in [-\Lambda, -L]\cr
e^{{\pi\over4}g^2(L+x)}\;,\qquad & x\in [-L, t-L]\cr
e^{{\pi\over2}g^2t}\;,\qquad & x\in [t-L, L]\cr
e^{{\pi\over4}g^2(L+t-x)}\;,\qquad & x\in [L, t+L]\cr
1\;,\qquad & x\in [t+L, \Lambda]}\right.\eqno(Gxtl)
$$
As long as $L$ and $t$ are both sufficiently large, the differences
between the two cases \(Gxt) and \(Gxtl) turn out to be insignificant.

Applying another quasi-diagonal approximation in combining \(alpham), \(CD),
and \(DCinv), we obtain the boundary values
$$\eqalign{
\alpha\-m(t) &\simeq e^{-i{|m|\pi\over\Lambda}t} \b{m} \sum_k
{\alpha^i\-k\over\b{k}} \lint {dx\over2\Lambda} {e^{i(m-k){\pi x\over\Lambda}}
(1- e^{-\beta{|k|\pi\over\Lambda}})\over
G\-k(x,t) - {e^{-\beta{|k|\pi\over\Lambda}}\over G\-k(x,t)}}\cr
&\;\;+ \b{m}\sum_k {\alpha'^f\-k\over\b{k}}
\lint {dx\over2\Lambda} {e^{i(k-m){\pi x\over\Lambda}}
 e^{-\beta{|k|\pi\over\Lambda}} \bigl( G\-k(x,t) - {1\over
G\-k(x,t)}\bigr)\over
G\-k(x,t) - {e^{-\beta{|k|\pi\over\Lambda}}\over G\-k(x,t)}}\cr
\alpha'\-m(0) &\simeq e^{i{|m|\pi\over\Lambda}t} \b{m} \sum_k
{\alpha'^f\-k\over\b{k}} \lint {dx\over2\Lambda} {e^{i(k-m){\pi x\over\Lambda}}
(1- e^{-\beta{|k|\pi\over\Lambda}})\over
G\-k(x,t) - {e^{-\beta{|k|\pi\over\Lambda}}\over G\-k(x,t)}}\cr
&\;\;+ \b{m}\sum_k {\alpha^i\-k\over\b{k}}
\lint {dx\over2\Lambda} {e^{i(m-k){\pi x\over\Lambda}}
\bigl(G\-k(x,t) - {1\over G\-k(x,t)}\bigr)\over
G\-k(x,t) - {e^{-\beta{|k|\pi\over\Lambda}}\over G\-k(x,t)}}\;.}\eqno(alphab)
$$
In a similar manner, we also obtain
$$\eqalign{
\alphabar\-m(0) &\simeq e^{-i{|m|\pi\over\Lambda}t} \b{m} \sum_k
{\alphabar^f\-k\over\b{k}}
\lint {dx\over2\Lambda} {e^{i(m-k){\pi x\over\Lambda}}
(1- e^{-\beta{|k|\pi\over\Lambda}})\over
G\-k(x,t) - {e^{-\beta{|k|\pi\over\Lambda}}\over G\-k(x,t)}}\cr
&\;\;+ \b{m}\sum_k {\alphabar'^i\-k\over\b{k}}
\lint {dx\over2\Lambda} {e^{i(k-m){\pi x\over\Lambda}}
\bigl(G\-k(x,t) - {1\over G\-k(x,t)}\bigr)\over
G\-k(x,t) - {e^{-\beta{|k|\pi\over\Lambda}}\over G\-k(x,t)}}\cr
\alphabar'\-m(t) &\simeq e^{i{|m|\pi\over\Lambda}t} \b{m} \sum_k
{\alphabar'^i\-k\over\b{k}}
\lint {dx\over2\Lambda} {e^{i(k-m){\pi x\over\Lambda}}
(1- e^{-\beta{|k|\pi\over\Lambda}})\over
G\-k(x,t) - {e^{-\beta{|k|\pi\over\Lambda}}\over G\-k(x,t)}}\cr
&\;\;+ \b{m}\sum_k {\alphabar^f\-k\over\b{k}}
\lint {dx\over2\Lambda} {e^{i(m-k){\pi x\over\Lambda}}
 e^{-\beta{|k|\pi\over\Lambda}} \bigl( G\-k(x,t) - {1\over
G\-k(x,t)}\bigr)\over
G\-k(x,t) - {e^{-\beta{|k|\pi\over\Lambda}}\over G\-k(x,t)}}\;.}
\eqno(alphabarb)
$$

Using \(alphab) and \(alphabarb) in \(sB) and \(JBB), we find the density
matrix
propagator ${\s J}$ to be given by
$$\eqalign{
{\s J}[\alpha^i\-k, \alphabar'^i\-m, &\alphabar^f\-m,\alpha'^f\-k;t]
 =\cr
&{1\over Z} \exp
\sum_{l,m}\bigl( \alphabar^f\-l \alpha^i\-m T\-{lm}
+ \alphabar'^i\-l \alpha'^f\-m T'\-{lm}
+ \alphabar^f\-l \alpha'^f\-m E\-{lm} + \alphabar'^i\-l \alpha^i\-m
A\-{lm}\bigr)\;,} \eqno(BBB)
$$
where
$$\eqalign{
T\-{lm} &\simeq e^{-i{|l|\pi\over\Lambda}t}
\lint {dx\over2\Lambda} {e^{i(l-m){\pi x\over\Lambda}}
(1- e^{-\beta{|m|\pi\over\Lambda}})\over
G\-m(x,t) - {e^{-\beta{|m|\pi\over\Lambda}}\over G\-m(x,t)}}\cr
T'\-{lm} &\simeq e^{i{|m|\pi\over\Lambda}t}
\lint {dx\over2\Lambda} {e^{i(l-m){\pi x\over\Lambda}}
(1- e^{-\beta{|m|\pi\over\Lambda}})\over
G\-m(x,t) - {e^{-\beta{|m|\pi\over\Lambda}}\over G\-m(x,t)}}\cr
E\-{lm} &\simeq
\lint {dx\over2\Lambda} {e^{i(m-l){\pi x\over\Lambda}}
 e^{-\beta{|m|\pi\over\Lambda}} \bigl( G\-m(x,t) - {1\over
G\-m(x,t)}\bigr)\over
G\-m(x,t) - {e^{-\beta{|m|\pi\over\Lambda}}\over G\-m(x,t)}}\cr
A\-{lm} &\simeq
\lint {dx\over2\Lambda} {e^{i(l-m){\pi x\over\Lambda}}
\bigl(G\-{m}(x,t) - {1\over G\-{m}(x,t)}\bigr)\over
G\-{m}(x,t) - {e^{-\beta{|m|\pi\over\Lambda}}\over G\-{m}(x,t)}}\;.}\eqno(TEA)
$$
These matrices are obtained using one new kind of quasi-diagonal approximation,
which may be illustrated by writing
$$\eqalign{
T\-{lm} &\simeq e^{-i{l\pi\over\Lambda}t\sgn{m}}
\lint {dx\over2\Lambda} {e^{i(l-m){\pi x\over\Lambda}}
(1- e^{-\beta{|m|\pi\over\Lambda}})\over
G\-m(x,t) - {e^{-\beta{|m|\pi\over\Lambda}}\over G\-m(x,t)}}\cr
&=e^{-i{|m|\pi\over\Lambda}t}
\lint {dx\over2\Lambda} {e^{i(m-l){\pi x\over\Lambda}}
(1- e^{-\beta{|m|\pi\over\Lambda}})\over
G\-m(x,t) - {e^{-\beta{|m|\pi\over\Lambda}}\over G\-m(x,t)}}\cr
&\simeq e^{-i{|m|\pi\over\Lambda}t}
\lint {dx\over2\Lambda} {e^{i(m-l){\pi x\over\Lambda}}
(1- e^{-\beta{|l|\pi\over\Lambda}})\over
G\-l(x,t) - {e^{-\beta{|l|\pi\over\Lambda}}\over G\-l(x,t)}}=T\-{ml}
\;,}\eqno(Ttrans)
$$
by substituting $x \to x - (x-t\sgn{m})$ and observing that $G\-m(- x +
t\sgn{m}) = G\-m(x,t)$.  (The last step in \(Ttrans) invokes the kind of
quasi-diagonal approximation already introduced in \(dinv).)

Equations \(BBB) and \(TEA)  provide us with the propagator for the density
matrix of the modes above $\Omega$ of the massless
scalar field, in the presence of a black body at inverse temperature $\beta$,
initially uncorrelated with the field.  The integrals in \(TEA) do not seem to
have closed form evaluations, but since $T$, $T'$, $A$, and $E$ are all
matrices formed by adding and multiplying matrices which are quasi-diagonal,
they must themselves also be quasi-diagonal.  (The inverse of a quasi-diagonal
matrix is quasi-diagonal; one way of showing that this statement applies in our
case is to consider the
massless field as the continuum limit of a lattice, in which case all the
matrices become finite dimensional, and the proposition is then elementary.
One might be concerned that the
diagonal entries of two quasi-diagonal matrices could cancel when they are
added, but the diagonal elements of the matrices in \(TEA) can be computed, and
they are not of order $g^2\over\Omega$.)  The quasi-diagonality of the matrices
in \(TEA) will permit us to interpret the black body propagator ${\s J}$, by
calculating some significant quantities to zeroth order in $g^2$.

\subhead{\bf C. Physical interpretation of the propagator}

For convenience in explaining the significance of \(BBB), we now give names to
some important spatial regions; these regions are illustrated in Figure 1.
The overlapping intervals $[-(L+t),L]$ and $[-L,(L+t)]$
will be referred to as the left and right {\it shadows} of the black body.  We
define the penetration depth $\lambda$ to be the smallest length for
which $e^{-{\pi\over2}g^4\lambda}$ is negligibly small.  The shadow regions
$[-(L+t),-(L+t-\lambda)]$ and $[(L+t-\lambda),(L+t)]$ are then called the {\it
penumbra}, while $[-(L+t-\lambda),-L]$ and $[L,(L+t-\lambda)]$ compose the
{\it umbra}.  The edge regions $[-L,-(L-\lambda)]$ and
$[(L-\lambda),L]$ are referred to as the {\it skin} of the black body, while
$[-(L-\lambda),(L-\lambda)]$ forms the {\it core}.  As long as both $t$ and $L$
are larger than $\lambda$, the differences between the two forms of $G(x,t)$
prescribed by \(Gxt) and \(Gxtl) have no effect on our identification of these
regions, since these differences are clearly insignificant in \(TEA).

With these relevant regions in mind, a simple interpretation of the propagator
is immediately suggested by the forms
of the four matrices in \(TEA).  We observe that $G\-k(x,t) - {1\over
G\-k(x,t)}$ vanishes for $x$ outside the shadow of the black body; it has
support within the left shadow for $k<0$, and within the right shadow for
$k>0$.  $E\-{mk}$ therefore seems to refer to thermal emission outwards from
the black body, and $A\-{mk}$ seems related to absorption of quanta that pass
into it.  As well, $(1-e^{-\beta{|k|\pi\over\Lambda}})
[G\-k(x,t) - {e^{-\beta{|k|\pi\over\Lambda}}\over
G\-k(x,t)}]^{-1}$ effectively vanishes within the umbral and black body zones,
but equals unity outside
the shadow.  $T\-{mk}$ and $T'\-{mk}$ are therefore suggestive of ordinary,
free field propagation in the regions causally disconnected from the black
body.

These impressions can be confirmed by some straightforward calculations
involving ${\s J}$.  Firstly, we can consider the final state into which the
vacuum evolves under ${\s J}$, and show that all $N$-point correlation
functions for this
state are the same as those for uniform, unidirectional thermal radiation, as
long as all the $N$ points are within the umbra of the same side of the black
body's shadow.  Secondly, we can show that the uniform thermal state at the
temperature of the black body evolves under ${\s J}$ into itself.  Finally, we
can set the temperature of the black body to zero, and obtain the ``quantum
wall effect'' whereby quanta are absorbed by the black body, without
reflection.  These three special cases suffice to illustrate that \(BBB) is
indeed the correct quantum mechanical description of the effect of a black body
on a massless field.

The final state evolving from initial vacuum is found simply by setting
$\alpha^i\-m = \alphabar'^i\-m = 0$ in \(BBB):
$$
\rho\-0[\alphabar\-m,\alpha\-k;t] = {1\over Z} \exp \sum_{km}
\alphabar\-m \alpha\-k E\-{mk}\;.\eqno(rhovac)
$$

The correlation functions for products of operators $\hat\phi(x\-1) ...
\hat\phi(x\-N)\hat\Pi(y\-1) ... \hat\Pi(y\-M)$ may be formed from the
expectation values
$$
\Big\langle \prod_{n=1}^N\left( \sum_m e^{-i{m\pi\over\Lambda}x\-n} {\hat
a^\dagger\-m u\-{mn}\over2\Lambda}\right)
\prod_{n=1}^M\left( \sum_m e^{i{m\pi\over\Lambda}y\-n} {\hat a\-m v\-{mn}\over
2\Lambda} \right) \Big\rangle \equiv \langle \hat{\s O}\rangle\;,\eqno(hatO)
$$
where $u\-{mn}, v\-{mn}$ are the factors $({|m|\pi\over\Lambda})^{\pm\12}$
applicable to either $\hat\phi$ or $\hat\Pi$.  This expectation value may be
computed by twice completing ``squares''\footnote{$^\dagger$}
       {Actually, bilinears in Bargmann-Fock variables are not squares, since
       $\alphabar$ and $\alpha$ are not complex conjugates, but the procedure
is
       exactly analogous to the usual method of manipulating quadratic
       exponents.}
in a Gaussian integral:
$$\eqalign{
\langle \hat{\s O}\rangle &= \int\!\prod_j {d\alpha\-j d\alphabar\-j d\beta\-j
d\bar\beta\-j\over (2\pi i)^2}\, e^{\sum_l (\alphabar\-l\beta\-l +
\bar\beta\-l\alpha\-l)} \rho_0[\alphabar\-l,\alpha\-m;t]
\langle\bar\beta\-j|\hat{\s O}|\beta\-j\rangle\cr
&=\left[\left(\prod_{n=1}^N {d\ \over dK\-n}\right)\left(\prod_{n=1}^M {d\
\over d\bar K\-n}\right) \exp {1\over2\Lambda}\sum_{n=1}^N \sum_{m=1}^M
\bar K\-m {\s U}\-{mn} K\-n \right]_{K\-n=\bar K\-n = 0}
\;,}\eqno(Xpec)
$$
where
$$\eqalign{
{\s U}\-{mn} &= \sum_{kl}
e^{-i{k\pi\over\Lambda}x\-n}u\-{kn}[(1-E)^{-1}-1]\-{kl}
e^{i{l\pi\over\Lambda}y\-m}v\-{lm} \cr
&\simeq \sum_{k}
e^{i{k\pi\over\Lambda}(y\-m-x\-n)}v\-{km}u\-{kn}\left[1-{1\over
G^2\-k(x\-n,t)}\right] {e^{-\beta{|k|\pi\over\Lambda}}\over
1- e^{-\beta{|k|\pi\over\Lambda}}}\;.}\eqno(UUU)
$$
We derive this by using quasi-diagonality to write
$$
(1-E)^{-1}\-{kl}\simeq \lint{dx\over2\Lambda}\,e^{i(k-l){\pi
x\over\Lambda}}{G\-l(x,t)-{e^{-\beta{|l|\pi\over\Lambda}}\over G\-l(x,t)}
\over G\-l(x,t) \left(1-e^{-\beta{|l|\pi\over\Lambda}}\right)}\;.\eqno(oneEinv)
$$

The expectation value \(Xpec) can be computed similarly (and trivially) for
states of uniform, isotropic thermal radiation, as well as for uniform thermal
radiation travelling in one direction only.  These states imply ${\s U}$
kernels analogous to that of \(UUU), which are
$$\eqalign{
{\s U}\-{mn} &= \sum_{k=-\infty}^\infty e^{i{k\pi\over\Lambda}(y\-m-x\-n)}
{1\over e^{\beta{|k|\pi\over\Lambda}}-1}\qquad\qquad\hbox{uniform,
isotropic}\cr
{\s U}\-{mn} &= \sum_{k=-\infty}^0 e^{i{k\pi\over\Lambda}(y\-m-x\-n)}
{1\over e^{\beta{|k|\pi\over\Lambda}}-1}\qquad\qquad
                     \hbox{uniform, left-directed}\cr
{\s U}\-{mn} &= \sum_{k=0}^\infty e^{i{k\pi\over\Lambda}(y\-m-x\-n)}
{1\over e^{\beta{|k|\pi\over\Lambda}}-1}\qquad\qquad\hbox{uniform,
right-directed}\;.}\eqno(compuuu)
$$
By comparing \(UUU) with \(compuuu) and referring to \(Gxt),
we can therefore see that, as far as measurements made outside the shadow of
the black body are concerned, the state represented by
$\rho\-0[\alphabar\-l,\alpha\-m;t]$ is indistinguishable from the vacuum; with
measurements made in the left (right) umbra, it is indistinguishable from the
state where only left-moving (right-moving) modes are thermally populated; and
for measurements inside the core of the body, it is indistinguishable from
isotropic thermal radiation.
(The asymmetry in \(Xpec) between the $x\-n$ and $y\-n$ is actually spurious,
since $G\-l(x,t)$ differs appreciably from $G\-l(y,t)$ only if $|x-y|\sim
{1\over g^2}$, but the thermal kernel defined by the sum over $\pm
{l\pi\over\Lambda} > \Omega$ decays at
least as fast as $1\over (x-y)$.  The asymmetry therefore contributes terms of
order $g^2$, which must be disregarded to be consistent in our ideal black body
approximation.)

For our second illustration, we compute the final state which evolves from
the thermal state at the black body temperature.  This initial state has
density operator
$$
\rho\-\beta[\bar\xi\-m,\xi\-m;0] = {1\over Z} \exp \sum_m
e^{-\beta{|m|\pi\over\Lambda}} \bar\xi\-m \xi\-m\;, \eqno(rhobeta)
$$
and using \(sJ), integrating by completing the ``square'' twice, it evolves
into
$$
\rho\-\beta[\alphabar\-l,\alpha\-m;t] =
{1\over Z} \exp \sum_{lm} \alphabar\-l [T \cdot (e^{\beta{|m|\pi\over\Lambda}}
-
A)^{-1} \cdot T'  +  E]\-{lm} \alpha\-m \;.\eqno(rhobetat)
$$
We can approximate
$$
(e^{\beta{|m|\pi\over\Lambda}}-A)^{-1}\-{lm} \simeq \lint{dx\over2\Lambda}\,
e^{i(l-m){\pi x\over\Lambda}} {G\-m(x,t) - {e^{-\beta{|m|\pi\over\Lambda}}\over
G\-m(x,t)}\over G\-m(x,t) (e^{\beta{|m|\pi\over\Lambda}}-1)}\;,\eqno(ebetamA)
$$
which leads to
$$\eqalign{
[T \cdot (e^{\beta{|m|\pi\over\Lambda}} &- A)^{-1}\cdot T']\-{lm}\cr
&\simeq  e^{i(m-l)\,\sgn{m}\,{\pi t \over\Lambda}}
\lint{dx\over2\Lambda}\, e^{i(l-m){\pi x\over\Lambda}}
{e^{-\beta{|m|\pi\over\Lambda}} (1-e^{-\beta{|m|\pi\over\Lambda}})\over
G^2\-m(x,t) - e^{-\beta{|m|\pi\over\Lambda}}}\cr
&=\lint{dx\over2\Lambda}\, e^{i(m-l){\pi x\over\Lambda}}
{e^{-\beta{|m|\pi\over\Lambda}} (1-e^{-\beta{|m|\pi\over\Lambda}})\over
G^2\-m(x,t) - e^{-\beta{|m|\pi\over\Lambda}}}\;.}\eqno(matrix)
$$
We make this last step by again changing variables $x\to -(x-t\sgn{m})$
(wrapping around periodically in the domain $[-\Lambda,\Lambda]$).
It immediately follows that
$$
[T \cdot (e^{\beta{|m|\pi\over\Lambda}} - A)^{-1}\cdot T' + E]\-{lm}
\simeq e^{-\beta{|m|\pi\over\Lambda}} \delta\-{lm}\;,\eqno(equil)
$$
which establishes the fact that
$$
\rho\-\beta[\alphabar\-m,\alpha\-m;t] = \rho\-\beta[\alphabar\-m,\alpha\-m;0]
+{\s O}({g^2\over\Omega})\;,
$$
as far as modes above the IR cut-off $\Omega$ are concerned.
The canonical ensemble at the temperature of
the ideal black body is indeed the equilibrium state.

Finally, we can consider the evolution of a one-particle
initial pure state, in the non-emitting case where $\beta\to\infty$,
so that $E\-{lm}\to 0$.  Changing to the $N$-particle basis by extracting the
co-efficients of
$\alphabar'^i\-m\alpha^i\-n$ and $\alphabar^f\-m\alpha'^f\-n$ in
${\s J}\vert_{\beta\to\infty}$, we find that
the final state evolving from the initial state
$$
\hat\rho\-i = \sum_{lm} \psi^i\-l\psi^{f*}\-m \big|{l\pi\over\Lambda}\big
\rangle\big\langle{m\pi\over\Lambda}\big|\;,
$$
where $|{l\pi\over\Lambda}\rangle$ denotes the one particle state with momentum
${l\pi\over\Lambda}$, has the density operator
$$
\hat\rho\-f = F |0\rangle\langle 0| + (1-F)\sum_{l,m}\psi^f\-l\psi^{f*}\-m
\big|{l\pi\over\Lambda}\big\rangle\big\langle{m\pi\over\Lambda}\big|\;,
\eqno(rhof)
$$
where $|0\rangle$ denotes the vacuum and
$$\eqalign{
F&\simeq {1\over Z} \sum_{lm} \psi^{i*}\-l A\-{lm} \psi^i\-m\cr
\psi^f\-l&\simeq {1\over \sqrt{1-F}} \sum_{m} T\-{lm}\psi^i\-m\cr
\psi^{f*}\-m&\simeq {1\over \sqrt{1-F}} \sum_l
\psi^{i*}\-l T'\-{lm}\;.}\eqno(evo)
$$
In the limit $\beta\to\infty$, we have
$$
T\-{lm}\big\vert_{\beta\to\infty} \simeq e^{-i{|l|\pi\over\Lambda}t}
\lint{dx\over2\Lambda}\, {e^{i(m-l){\pi x\over\Lambda}}\over G\-{-l}(x,t)}\;.
\eqno(Tlm)
$$
This ensures that an initial state localized within the umbra on
the left (right) side of the black body evolves into a state in which
right-moving (left-moving) modes above the IR cut-off are negligibly excited.
A state localized within the core of the black body evolves into one in which
no modes above the cut-off are significantly excited.  This then implies that
the probability of a signal passing through the black body is negligible.
Note that, when $\beta\to\infty$, we have $(A+ T\cdot T')\-{lm} \simeq
\delta\-{lm}$. This shows that the total probability is conserved, while the
probability of a particle being present decreases from one.

Our model and the quasi-diagonal approximation we have used to analyse it
therefore exhibit absorption, thermodynamic equilibrium, and black body
radiation.  We have presented, in an idealized model in one spatial  dimension,
a description from quantum mechanical first principles of the interaction of a
macroscopic black body with a massless quantum field, whose behaviour must
closely resemble that of light.

\head{\bf IV. Discussion}

\subhead{\bf A. Summary}

The use of influence functionals to describe the second order effects of an
unobserved environment has not received its due emphasis in texts on quantum
theory.  Among second order effects, there can be some that are enhanced by
large parameters that do not appear at lower orders.  These are very
important, since they are apparently responsible for the significant phenomena
of quantum measurement and thermal dissipation.   As discussed by Feynman and
Vernon in their classic paper introducing influence
functionals[\cite{FV}],
independent harmonic oscillators can be used to describe any environment, as
long as we are only interested in effects up to second order in the coupling
between environment and observed system.  To this
order, a general quantum mechanical environment having allowed transitions with
energy differences ${\omega\-k}$ is equivalent to a set of harmonic oscillators
with those natural frequencies, with effective coupling strengths proportional
to the matrix elements of the interaction Hamiltonian\footnote{$^\dagger$}
{One might be concerned that, while the dimensionful $g^2$ set the absorption
scale $\lambda$ in our one-dimensional calculation, in actual spacetime the
electromagnetic fine structure constant $\alpha\-{em}$ is dimensionless.  In
fact, though, the effective couplings which appear when realistic matter is
approximated by harmonic oscillators must be proportional to dipole moments of
molecules, etc., and so will still have the correct
dimensionality.}[\cite{FV}].

This result is derived in the Appendix.  To first order in the coupling,  a
generic quantum environment may be treated as a classical source; the first
improvement beyond this treatment, valid to second order in the coupling, is to
model a generic environment as a collection of independent oscillators.  Our
treatment of matter as a collection of baths of harmonic oscillators is
therefore not merely a gross idealization, but is actually accurate up to
second order in perturbation theory.

{}From another viewpoint, though, our model of a free quantum field interacting
with first-quantized matter can actually be considered as
a way of approaching some non-perturbative physics.  The atoms and molecules
represented by heat baths are in reality bound states involving the
electromagnetic field.  Interacting fields can effectively have many more
degrees of freedom than free fields, because their Hilbert spaces include
arbitrary numbers of bound states.  These bound states couple to the unbound
modes, and can have significant effects if they are present in macroscopic
quantities --- as in a black body of macroscopic size.

That these effects can be significant despite weak coupling can be starkly
demonstrated by calculating the energy radiated by a black body.  From our one
dimensional model in Section III, we find (in the limit where the infrared
cutoff is removed) that the outgoing radiation carries
energy away at the rate
$$
{\s R} = {\pi^2\over6\beta^2} = {\pi^2\over6\hbar k\-B^2}T^2\eqno(radrate),
$$
making $\hbar$ and the Boltzmann constant $k\-B$ explicit;
$T$ is the black body temperature.
This is the one-dimensional, spin zero version of the Stefan-Boltzmann law,
$$
{\s R}\-B = \sigma T^4\;.\eqno(Stefan)
$$
The Stefan-Boltzmann constant $\sigma$ is independent of the electromagnetic
coupling; yet it governs electromagnetic radiation generated by matter.  From
the analysis of Section III we can now understand this seeming paradox: because
thermal radiation occurs in situations with microscopically large time and
distance scales, one must perturb {\it separately} in ${g^2\over\Omega}$ and
$e^{-{\pi\over2}g^2\lambda}$.  The limit where both these quantities are small
is qualitatively different from the limit where $g^2$ is set exactly to zero.
The Stefan-Boltzmann constant might therefore be said to contain an
``invisible'' charge-dependent  factor of a form like
$(1-e^{-{\pi\over2}g^2\lambda})$, which would vanish if the  electromagnetic
coupling were literally set to zero, but which is negligibly different from
unity for macroscopic black bodies.

The condition that the black body has negligible  reflectivity implies that the
spatial growth and decay of the field solutions must be very small compared to
their oscillatory frequency.  This in turn implies the validity of the
quasi-diagonal approximation, whereby modes of a given frequency evolve
exclusively into modes of the same frequency, but with spatial  modulation that
becomes significant over large enough distances.  An approximation of this kind
is implicitly being invoked whenever we speak of light of definite frequency
and direction being restricted to finite regions of space.   The prevalence in
nature of the weak effective couplings that justify such concepts is ultimately
due to the  smallness of the QED coupling\footnote{$^\dagger$} {The effective
coupling may be even weaker,  if small matrix elements are involved in deriving
the independent oscillator  heat bath via second order perturbation theory.
The density of degrees of freedom in material media is also a factor.  It
does not seem unreasonable to conclude that quantum mechanics is, in its own
way, as medium-dependent as classical optics.}.    The vanishing reflectivity
of a black body can thus be thought of as a mere indicator of the underlying
physics of weak coupling.

Compared to the physics of weak coupling, the enhancement of second order
effects by macroscopically large parameters such as our $\lambda$ is certainly
a less familiar physical phenomenon.  Yet it also permits a powerful type of
approximation, elucidating ``qualitative'' properties which, like black body
radiation and quantum measurement, have traditionally been treated by axiom
rather than by perturbation.  The opacity of a black body can be considered as
an indicator that a ``large parameter'' approximation is appropriate.

With two very different, very drastic approximations applying at once,
it is little wonder that a non-reflective, opaque object is such a clean
theoretical subject, and why its behaviour is so universal (depending only on
temperature).  The highly non-trivial implications of total absorptivity, first
deduced from thermodynamic axioms, have now been traced to their microscopic
origins in quantum theory. The results of our seemingly very idealized model,
with uncountably many independent harmonic oscillators coupled to the field
at every point, are thus quite universal.  Is this model then more realistic
than it at first seems?

\subhead{\bf B. Justifying the model}

Apart from the reduction to one dimension, this model is in fact quite
realistic.  In a disordered collection of many molecules there are very many
allowed transition energies, and the continuous spectrum $I(\omega)$  becomes a
reasonable approximation to a discrete spectrum of equivalent oscillators,
that will not break down until one considers a time scale on the order of the
inverse of the spectral spacing.  Since a continuous spectrum of harmonic
oscillators has infinite heat capacity, one would expect this time scale to be
associated with heating and cooling of the black body.  These important
phenomena are not described by our model, because of our assumption of a
continuous spectrum.

In Section III, however, we assumed that there is an independent heat bath
interacting with the field at every point in the black body, so that   our
model is continuously dense in space, as well as in frequency.  This is a valid
approximation for a discrete model, as long as we consider wavelengths of light
long compared with the lattice spacing.  We are therefore certainly only
modeling the effect of matter on light with wavelengths above  the X-ray range.
In fact, we are assuming that each heat bath represents a grain or clump of
molecules or atoms that effectively couples to the field at a single point, and
so must be  of a size less than the wavelengths of interest.  For wavelengths
as small as the near ultraviolet range, this permits us to have on the order of
a thousand  atoms per heat bath.  This number of degrees of freedom must still
suffice to  give an individual heat bath cooling time scale much longer than
our relaxation time scale $g^{-2}$.  Is there a conflict between our
assumptions of continuity in space and frequency?

We can estimate the cooling time scale for a thousand-atom grain by dividing
its classical thermal energy ${3\over2}Nk\-B T$ by the rate of energy loss of a
spherical, classical black body of radius two to three nanometers.  Using
the Stefan-Boltzman law \(Stefan), we find the cooling time to be
$$
\tau\-c \sim {1500 k\-B T\over 4\pi r^2 \sigma T^4} \sim 10^4 T^{-3} \ s\cdot
K^3\;.\eqno(tauc)
$$
We can bound $g^2$ phenomenologically by assuming that our IR cut-off is in the
far infrared, at around $10^{12}$ Hz, and by knowing that dark materials can be
opaque at a thickness of around $10^{-4}$ metres, giving a range for $g^{-2}$
of perhaps $10^{-10}$ to $10^{-6}$ seconds.  For temperatures ranging into the
thousands of Kelvin, therefore, we have $\tau\-c >> g^{-2}$.  The continuous
spectral density approximation should therefore not break
down for our local heat baths until the processes described in
Section III are well established.  At this  point issues arise, concerning heat
transport within the medium, that are beyond the scope of this paper.  Such
issues probably relate to a further limitation on our model: in considering
each heat bath  to be independent of all the others, we are explicitly and
qualitatively excluding any effects involving longer range correlations in
matter, such as conduction.  Typical weak dielectrics, on the other hand,
should be adequately represented by our model.

The implicit infrared and ultraviolet cutoffs in our model are not seriously in
conflict with the ohmic spectral density assumed in Section III.
Exact constancy of the spectral density is not essential for the results of
that Section, since slowly varying $I(\omega)$ leads to similar results with a
frequency dependent decay constant $\Gamma(\omega)$.  As long as
$e^{-2L\Gamma(\omega)}$ is negligible, the body is effectively uniformly
absorptive.

As far as infrared to near ultraviolet light is concerned, then, the black body
presented in this paper is actually a fairly accurate representation of a
medium composed of closely packed grains, whose size and packing distances are
on the order of a few nanometers, each of which consists of a few thousand
atoms.  Dense, fine-grained deposits of carbon atoms occur as lampblack,
which has long been used for making actual black bodies black
(although this is partly because it makes a rough surface --- a feature
we have not represented).  This paper thus turns out to be about soot.

\subhead{\bf C. Issues raised by the model}

The problem we have studied presents a new face of the basic quantum dilemma
regarding localization in position and momentum.  One may often think of a
black
body as thermally exciting outgoing field modes, and absorbing incident modes,
but this is misleading.  The positional and directional localization
involved in identifying outgoing and incoming radiation is more subtle than the
one-to-one mapping by which we associate Fourier modes with particle momenta.

This subtlety can be illustrated by sketching the extension of our work to a
black body which is a solid sphere in three dimensions.  Choosing spherical
polar co-ordinates, the radial eigenfunctions of the free field Hamiltonian
are spherical Bessel and Neumann functions.  While these can be combined into
spherical Hankel functions, describing incoming and outgoing waves, the Neumann
functions diverge at the origin, and the only physical modes are pure Bessels,
representing standing waves.  There are not enough field degrees of freedom,
then, to be decomposed into orthogonal outgoing and incoming modes.

The black body environment surrounding the origin changes the path integral
saddlepoint, and modifies the Euler-Lagrange equations.  Taking the zero
angular momentum mode as an example, we find
$f\-0(r) =  {e^{ikr \pm \Gamma r}\over r} - {e^{-ikr \mp \Gamma r}\over r}$
appearing in solutions, instead of ${\sin kr\over r}$.  Radial  functions that
are predominantly incoming or outgoing (outside the black body) will  therefore
appear in the three-dimensional analogue of \(TEA).  But the number of
orthogonal field modes is not suddenly doubled, and each of the incoming and
outgoing sets of eigenfunctions could in fact be expanded in terms of the
other. There is only a half-line $k\in [0,\infty]$ worth of radial degrees of
freedom to be described.

This point is one more illustration among many that the wave-particle duality
of quantum fields is not completely explained by referring to the expansion of
the field operator in plane waves.  While it is convenient for scattering
processes in negligible gravity, this notion of particles fails to be physical
in curved spaces or accelerating frames; it also requires elaboration (via our
quasi-diagonal approximation) to describe emission and absorption from
macroscopic bodies at rest in flat space. A really satisfactory formulation of
particle states would be of great value, and it has yet to be found.

A second issue is raised by our zero-temperature, purely absorptive limit.  The
``quantum wall effect'' blocks field excitations from passing through the black
body, but it does not affect the ground state of the field (as would the
reflecting boundary condition one typically assigns to a conducting plate).
This implies that there is no Casimir effect, at zeroth order in
${g^2\over\Omega}$ and   $e^{-{\pi\over2}g^2 L}$, on parallel plates of
non-conducting dielectric[\cite{Candelas}].  The case of a conducting body,
excluded in this paper, should also admit a similar kind of treatment, with the
independent point heat baths replaced by a gas of free electrons in a box of
size $2L$.  It would be interesting to study such a model in order to
investigate the Casimir  effect without simply imposing boundary conditions, as
well as to understand the important phenomenon of thermal radiation by hot
metal.

Finally, we could relax the assumption of continuous spectral density which
implied infinite heat capacity and discarded radiative cooling.  As the black
body cooled to low temperatures, we would expect that any information it had
absorbed from the field would be re-emitted, as it approached its unique ground
state.  This problem would also provide an interesting comparative model for
the evaporation of a black hole through the Hawking effect.  Although a black
hole is believed to undergo radiative heating rather than cooling, the issue of
the re-release of trapped information is common to both black systems.

\head{\bf Acknowledgements}

The author wishes to thank R.C. Myers, W.H. Zurek, J.P. Paz, and S. Habib for
valuable discussions, and to acknowledge the hospitality of Los Alamos National
Laboratory during part of this work.  This research was supported in part by
NSERC of Canada, and by a Carl Reinhardt Fellowship of McGill University.

\vfill

\head{\bf Appendix: Influence functional for a weakly coupled environment}

Consider an environment represented by the single degree of freedom $q$, with
conjugate momentum $p$.  Let this environment be coupled to an observed system,
such that the total Hamiltonian operator is
$$
\hat{H} = \hat{H}\-{S} + \hat{H}\-E + \epsilon\,\hat{A}(p,q)\,\hat{Q}\;.
\eqno(applint)
$$
The coupling constant $\epsilon$ is assumed to be small and dimensionless.
$\hat{H}\-{S}$ and $\hat{Q}$
are operators on the observed system, while $\hat{H}\-E$ and
$\hat{A}(\hat p,\hat q)$ are operators in the unobserved sector.

Let $|K\rangle$ denote eigenstates of the environmental Hamiltonian, such that
$\hat H\-E |K\rangle = E\-K |K\rangle$, assuming no special form for
$H\-E(p,q)$.   We can then in principle define the $q$ and $p$ representation
wave functions $\psi\-K(q)=\langle q|K\rangle$ and  $\Psi\-K(p)=\langle
p|K\rangle$.  Let the mixed initial state of the environment at time $s=0$ be
described by the density operator $\hat R = :\!\!R(\hat{p},\hat{q})\!\!:$,
where the colonss imply normal ordering by placing all $\hat p$'s to the left
of
all $\hat q$'s, so that $R(p,q)$ is the Wigner function
$\langle p|\hat R |q\rangle$.  $\hat R$ has the matrix elements
$$
\langle K|\hat R |L\rangle \equiv R\-{KL}
$$
in the basis of energy eigenstates.  The Wigner function may be expressed in
terms of these matrix elements using wave functions:
$$
R(p,q)=\sum_{K,L}\Psi\-K(p)R\-{KL}(0)\psi^*\-L(q)\;.\eqno(initR)
$$

For the environment with this Hamiltonian and initial state, the influence
phase $V[Q,Q']$ is given by the path integral
$$\eqalign{
e^{iV[Q,Q']} &= \int\!dp\-f dq'\-f dq\-i dp'\-i\,
R(q\-i,p'\-i)\, e^{-{i\over\hbar}p\-f q'\-f} \cr
\times&\int_{q\-i;0}^{p\-f;t}\!\D{p}\D{q}\!\int_{p'\-i;0}^{q'\-f;t}\!\D{p'}
\D{q'}\,
e^{{i\over\hbar}\int_0^t\!ds\,\bigl((p\dot{q} - p'\dot{q}') - (H\-E -
H'\-E) -\epsilon (A Q -A'Q') \bigr)} \;.}\eqno(Fpi)
$$
Here $A'$ and $H'\-E$ are short for $A(p',q')$ and $H\-E(p',q')$.

This path integral may be computed exactly in the harmonic case where
$H\-E(p,q) = \12(p^2 + \omega^2 q^2)$ and $A(p,q)=q$.  We find[\cite{FV}]
$$\eqalign{
V[Q,Q'] &\equiv
 {i\epsilon^2\over2\hbar \omega}\int_0^t\!ds\!\int_0^s\!ds'\,
(Q-Q')\-s \cr
&\qquad\times\bigl((Q-Q')\-{s'}\coth{\beta\hbar\omega\over2} \cos\omega(s-s')
 -i(Q+Q')\-{s'}\sin\omega(s-s')\bigr)\;,}\eqno(iho)
$$
where $s$ and $s'$ are time variables, $(Q-Q')\-s \equiv Q(s)-Q'(s)$, and $t$
is the final time.

The influence phases for any number of uncoupled and initially uncorrelated
environmental degrees of freedom simply add[\cite{FV}], and if a number of
uncoupled systems are in a thermal state, each will be separately in a
thermal state, uncorrelated with the other systems.  Consider, therefore, an
environment consisting of a number $N$ of independent harmonic oscillators,
possibly having different natural frequencies $\omega\-k$ and
coupling strengths $\epsilon\-k$.  If this environment is
initially in a thermal state and uncorrelated with the observed system, it will
have an influence functional of the form
$$\eqalign{
V[Q,Q'] &= {i\over\hbar^2} \int_0^\infty\!d\omega\, I(\omega)
\int_0^t\!ds\!\int_0^s\!ds'\,(Q-Q')\-s\cr
&\qquad\times \Bigl(\coth{\beta\omega\over2} (Q-Q')\-{s'}\cos\omega(s-s')
-i (Q+Q')\-{s'}\sin\omega(s-s')\Bigr)
   \;,}\eqno(ihosum)
$$
where
$$
I(\omega) = {\hbar\over2}\sum_{k=1}^N \delta(\omega-\omega\-k)
                                           {\epsilon^2\-k\over\omega\-k} \;.
\eqno(appIome)
$$

As we have seen in Section II, above, a coupling $A(p,q) = p$ leads to an
influence phase of the same form as \(ihosum).

Returning to the environment with only a single degree of freedom, we now
consider the case of general $H\-E(p,q)$ and $A(p,q)$.  As long as the
coupling $\epsilon$ is small enough, equation \(Fpi) may be
approximated perturbatively.  To first order in $\epsilon$, we have
$$\eqalign{
e^{iV[Q,Q']} &=\int\!dp\-f dq'\-f dq\-i dp'\-i\,
R(q\-i,p'\-i)\,e^{-{i\over\hbar}p\-f q'\-f}\cr
\times&\int_{q\-i;0}^{p\-f;t}\!\D{p}\D{q}\!
\int_{p'\-i;0}^{q'\-f;t}\!\D{p'}\D{q'}\,
e^{{i\over\hbar}\int_0^t\!ds\,\bigl((p\dot{q} - p'\dot{q}') -
(H\-E - H'\-E) \bigr)} \cr
&\qquad\qquad\times
\left(1 -{i\epsilon\over\hbar}\int_0^t\!ds\, (AQ-A'Q') +
{\s O}(\epsilon^2)\right)\cr
&= 1 - {i\epsilon\over\hbar}\int_0^t\!ds\, \sum_{L,M} R\-{LM} A\-{ML}
e^{-i\omega\-{LM}s} (Q-Q')\-s + {\s O}(\epsilon^2)\cr
&= \exp\Bigl(-{i\epsilon\over\hbar}\int_0^t\!ds\,
\langle A(s)\rangle (Q-Q')\-s\Bigr) + {\s O}(\epsilon^2)\;.}\eqno(ford)
$$
In \(ford) we use the following notation:
$$\eqalign{
A\-{LM} &\equiv \langle L|\hat A|M\rangle\cr
\omega\-{LM} &\equiv {E\-L-E\-M\over\hbar}\cr
\langle A(s)\rangle &\equiv \sum_{L,M} R\-{LM} A\-{ML}
e^{-i\omega\-{LM}s}\;.}
$$

When we recall that the influence phase appears in any path integral over
observed degrees of freedom as an additional term in the exponent,
$$
e^{{i\over\hbar}\bigl(S[Q] - S[Q']\bigr)}\to
e^{{i\over\hbar}\bigl(S[Q] - S[Q'] + V[Q,Q']\bigr)}\;,
$$
we can see that the first order term in \(ford) clearly constitutes a
perturbation to the observed sector action $S[Q]$, with the expectation value
$\langle A(s)\rangle$ acting as a classical external source.

Since the first order term leads to effects that are already well studied in
the regime of closed system quantum mechanics, we generally wish to concentrate
on the specifically open quantum behaviour produced by the higher order terms.
If we restrict our attention to environments that are approximately in
equilibrium, so that $R\-{MN} = {1\over Z} e^{-\beta E\-M} \delta\-{MN}$, and
assume that the expectation value of $\hat A$ vanishes\footnote{$^\dagger$}{By
considering a slightly generalized thermal state, we can in fact include
first order external sources together with the second order terms described in
this paper.}, then the first
non-trivial term in \(Fpi) is the second order contribution:
$$\eqalign{
e^{iV[Q,Q']} &=\int\!dp\-f dq'\-f dq\-i dp'\-i\,
R(q\-i,p'\-i)\,e^{-{i\over\hbar}p\-f q'\-f}\cr
&\qquad\times
\int_{q\-i;0}^{p\-f;t}\!\D{p}\D{q}\!\int_{p'\-i;0}^{q'\-f;t}\!\D{p'}\D{q'}
e^{{i\over\hbar}\int_0^t\!ds\,\bigl((p\dot{q} - p'\dot{q}') - (H\-E - H'\-E)
\bigr)}\cr
&\qquad\qquad\times
\left(1 -{\epsilon^2\over2\hbar^2}\left[\int_0^t\!ds\, (AQ-A'Q')\right]^2
+ {\s O}(\epsilon^3)\right)\cr
&=1-{\epsilon^2\over\hbar^2}{1\over Z}\int_0^t\!ds\!\int_0^s\!ds'\,\sum_{L,M}
e^{-\beta E\-L} A\-{LM} A\-{ML} (Q-Q')\-s \cr
&\qquad\qquad\times(Q\-{s'} e^{i\omega\-{LM}(s-s')} -
Q'\-{s'} e^{-i\omega\-{LM}(s-s')}) + {\s O}(\epsilon^3)\cr
&=1 -{\epsilon^2\over\hbar^2}{1\over Z}\int_0^t\!ds\!\int_0^s\!ds'\,\sum_{L>M}
(e^{-\beta E\-M} - e^{-\beta E\-L}) A\-{LM} A\-{ML} (Q-Q')\-s\cr
&\qquad\qquad\times \Bigl(
(Q-Q')\-{s'} \coth{\beta\hbar\omega\-{LM}\over2}\cos\omega\-{LM}(s-s')\cr
&\qquad\qquad\qquad\qquad
- i(Q+Q')\-{s'} \sin\omega\-{LM}(s-s')\Bigr) +{\s O}(\epsilon^3)\;.}\eqno(sord)
$$

This second order term cannot be represented by any modification of the action;
it contains qualitatively new behaviour characteristic of an open quantum
system.  Using the smallness of $\epsilon^2$, we can re-write \(sord) as
$$\eqalign{
V[Q,Q']& = {i\over\hbar^2} \int\!d\omega\, G(\omega;\beta)
\int_0^t\!ds\!\int_0^s\!ds'\,[q(s)-q'(s)]\cr
&\times \Bigl(\coth{\beta\omega\over2} [q(s')-q'(s')]\cos\omega(s-s')
-i [q(s')+q'(s')]\sin\omega(s-s')\Bigr)\;,}\eqno(Vsecord)
$$
where the effective spectral density is
$$
G(\omega;\beta) = {\epsilon^2}{1\over Z}\sum_{L>M} \delta(\omega-\omega\-{LM})
A\-{LM} A\-{ML} (e^{-\beta E\-M} - e^{-\beta E\-L})\;.\eqno(Gomebet)
$$

The restriction of the environment to a single degree of freedom is actually
irrelevant to the above derivation, and may be considered a mere notational
convenience.  Comparing \(Vsecord) and \(ihosum), therefore, we see that any
weakly coupled environment that is initially uncorrelated with the system and
in a thermal state is equivalent, at a fixed temperature, to a bath composed
of independent harmonic oscillators.  It is worth emphasizing that the
effective spectral density $G(\omega;\beta)$ can be temperature dependent, as
this fact seems to have escaped comment heretofore.

In fact, \(Vsecord)  can also apply in a situation where the weak coupling
approximation is not as straightforward, namely the case of a large number $N$
of independent (\ie uncoupled and initially uncorrelated) environmental degrees
of freedom, each of which is weakly coupled to the observed system, but not so
weakly that the total effect of the environment has to be small.  In this case
the Hamiltonian may be written as
$$
\hat{H} = \hat{H}\-{S} + \sum_{k=1}^N\left(
\hat{H}\-k +\epsilon\-k\,\hat{A}\-k(p\-k,q\-k)\,\hat{Q}\right)\;,\eqno(Nham)
$$
where all of the $\epsilon\-k$ are much less than unity.

Assuming again the initial thermal state, with vanishing expectation values
$\langle A\-{k}(s)\rangle$, we find
$$\eqalign{
e^{iV[Q,Q']} &=\int\!dp\-f dq'\-f dq\-i dp'\-i\, e^{-{i\over\hbar}p'\-f q\-f}
R(q\-i,p'\-i)\cr
\times&\int_{q\-i;0}^{p\-f;t}\D{p}\!\D{q}\!\int_{p'\-i;0}^{q'\-f;t}\!\D{p}
\D{q'}\cr
&\ \times\prod_{k=1}^N\,
e^{{i\over\hbar}\int_0^t\!ds\,
\bigl((p\-k\dot{q\-k} - p\-k'\dot{q\-k}') - (H\-k - H'\-k) \bigr)}\cr
&\qquad\times
\left(1 -{\epsilon^2\-k\over\hbar^2}\left[\int_0^t\!ds\,
(A\-k Q - A'\-k Q')\right]^2 + {\s O}(\epsilon^3\-k)\right)\cr
&=\prod_{k=1}^N\Bigl(
1-{\epsilon^2\-k\over\hbar^2}
\int_0^t\!ds\!\int_0^s\!ds'\,\int_0^\infty\!d\omega\, G\-k(\omega;\beta)\cr
&\qquad\qquad\times
\bigl((Q-Q')\-{s'} \coth{\beta\hbar\omega\over2}\cos\omega(s-s')\cr
&\qquad\qquad\qquad -
i(Q+Q')\-{s'} \sin\omega(s-s')\bigr) +{\s O}(\epsilon^3\-k)\Bigr)\;.}
\eqno(Nsord)
$$
Here $G\-k(\omega;\beta)$ is the effective spectral density for the $k$th
environmental degree of freedom, defined similarly to $G(\omega;\beta)$ in
\(Gomebet).

In the case of large $N$, we can now now approximate \(Nsord) by perturbing in
$1/N$. The  leading order term may be found by defining $\epsilon^2\-k =
{\eta\-k\over N}$ for finite $\eta\-k$, and taking  the limit $N\to\infty$.  In
this limit, the product
$$
\lim_{N\to\infty} \prod_{k=1}^N \Bigl(1 - {\eta\-k\over N}f\-k\Bigr) =
\exp{-\sum_{k=1}^N \eta\-k f\-k}\;,
$$
and so we see that the influence phase for the environment with a large number
of weakly coupled degrees of freedom is given by
$$\eqalign{
V[Q,Q']
&= \bigl(1+{\s O}(\epsilon\-k) + {\s O}({1\over N})\bigr)\times
{i\over\hbar^2}
\int_0^t\!ds\!\int_0^s\!ds'\,\int_0^\infty\!d\omega\,
{\s G}(\omega;\beta)\cr
&\qquad\qquad\times
\bigl((Q-Q')\-{s'} \coth{\beta\hbar\omega\over2}\cos\omega(s-s')\cr
&\qquad\qquad\qquad\qquad -
i(Q+Q')\-{s'} \sin\omega(s-s')\bigr)\;,}\eqno(largewk)
$$
where the effective spectral density is
$$
{\s G}(\omega;\beta) = \sum_{k=1}^N \epsilon^2\-k
G\-k(\omega;\beta)\;.\eqno(sGomebet)
$$
Equation \(largewk) is, to leading order in $\epsilon\-k$ and $1/N$, of
exactly the same form as \(Vsecord) and \(ihosum).

For large enough $N$, quantities such as $\int_0^\infty\!d\omega\, {\s
G}(\omega;\beta)$ --- a crucial term in derivations of wave function
collapse[\cite{wfcollapse}] --- may be extremely large, even though the
$\epsilon\-k$ are all  small. It is important to note, therefore, that even
though the coupling of the observed system to each environmental degree of
freedom may be weak, there may be so many such degrees of freedom that their
effect on the system can be large.  Equation \(largewk) shows that the
approximation of a generic  environment by a bath of independent oscillators
does not break down in this case.  We have therefore confirmed that a bath of
independent harmonic oscillators provides an adequate model for a weakly
coupled but macroscopically large environment, capable of inducing  drastic
effects like quantum measurement.

\vfill

\references

\singlespace

\refis{Jacobsen} Theodore Jacobson, \pr D48, 728, 1993, and references
therein.

\refis{deco} R. Omn\`es, \journal Rev. Mod. Phys., 64, 339, 1992; J.J.
Halliwell, \pr D46, 1610, 1992;  M. Gell-Mann and J.B. Hartle, \pr D47, 3345,
1993.

\refis{deco2} J.P. Paz and W.H. Zurek, \pr D48, 2728, 1993.

\refis{wfcollapse} W.G. Unruh and W.H. Zurek, \pr D40, 1071, 1989; B.L. Hu,
Juan Pablo Paz, and Yuhong Zhang, \pr D45, 2843, 1992.

\refis{Grabert} H. Grabert, P. Schramm, and G.-L. Ingold, \journal Phys. Rep.,
168, 115, 1988.

\refis{FV} R.P. Feynman and F.L. Vernon, \journal Ann. Phys. (N.Y.), 24, 118,
1963.

\refis{DvH} L. van Hove, \journal Physica, 21, 517, 1955; E.B. Davies, \journal
Commun. Math. Phys., 39, 91, 1974.

\refis{CaldLegg} A.O. Caldeira and A.J. Leggett, \pr A31, 1059, 1985.

\refis{ItZub} Claude Itzykson and Jean-Bernard Zuber, {\it Quantum Field
Theory} (McGraw-Hill; New York, 1980), p. 435 ff.

\refis{Planck} M. Planck, \journal Verh. Deutsch. Phys. Ges., 2, 237, 1900.

\refis{RayleighJeans} J.W.S. Rayleigh, \journal Phil. Mag., 49, 539, 1900;
J.H. Jeans, \journal Phil. Mag., 10, 91, 1905.

\refis{atten} B.R. Mollow, \pr 168, 1896, 1967.

\refis{Candelas} P. Candelas, \journal Ann. Phys., 143, 241, 1982.

\endreferences

\vfill\vfill
\endit